\newif\ifblind
\blindfalse    

\newif\ifarxiv
\arxivtrue    

\newif\iflinenumbers
\linenumbersfalse    

\ifarxiv
\documentclass[12pt]{article}
\usepackage{PRIMEarxiv}
\usepackage[font=small,skip=4pt]{caption} 

\else
\documentclass[12pt]{article}
\usepackage[font=small,skip=4pt]{caption} 
\usepackage[pagewise]{lineno}

\fi

\usepackage{xcolor}
\usepackage{fancybox}   

\usepackage{listings}
\usepackage{natbib}
\usepackage{comment}
\usepackage[utf8]{inputenc} 
\usepackage[T1]{fontenc}    
\usepackage{hyperref}       
\usepackage{url}            
\usepackage{booktabs}       
\usepackage{amsfonts}       
\usepackage{nicefrac}       
\usepackage{microtype}      
\usepackage{fancyhdr}       
\usepackage{graphicx}       
\usepackage{latexsym,multirow}
\usepackage{amssymb,amsmath,bm,mathtools,bbm}
\usepackage{arydshln}
\usepackage{theorem}
\usepackage{lmodern}
\usepackage{setspace}
\graphicspath{{media/}}     
\usepackage{cleveref} 
\usepackage{chngcntr}
\usepackage{apptools}
\usepackage{subcaption}     
\usepackage{booktabs}

\usepackage{algorithm}
\usepackage{algpseudocode}

\ifarxiv
\usepackage[]{geometry}
\else
\usepackage[margin=1in]{geometry}
\fi

\usepackage{tikz}
\usetikzlibrary{arrows, positioning}
\usepackage{pifont}

\newcommand\plim{\stackrel{\rm p}{\rightarrow}}

\newcommand\wmM{\widehat{\mathcal{M}}}
\newcommand{\mM}{{\mathcal{M}}}

\newcommand\E{\mathbb{E}}

\theoremstyle{plain}
\theoremheaderfont{\scshape}
\newtheorem{assumption}{Assumption}

\newtheorem{proof}{Proof}

\newtheorem{result}{Result}

\title{Large Language Models for Statistical Inference: \\ Context Augmentation with Applications to the Two-Sample Problem and Regression
}

\ifblind
  \author{} 
\else
\author{
  \textbf{Marc Ratkovic}\\
  Professor of Social Data Science \\
  Department of Political Science\\
  Department of Data Science\\
  University of Mannheim \\
  Mannheim, DE\\
  \texttt{marc.ratkovic@uni-mannheim.de} 
}
\fi

\begin{document}
{  \maketitle }
\iflinenumbers
\linenumbers
\fi

\begin{abstract}
 \setstretch{1.25}
 \singlespacing
 
We introduce context augmentation, a data-augmentation approach that uses large language models (LLMs) to generate contexts around observed strings as a means of facilitating valid frequentist inference.  These generated contexts serve to reintroduce uncertainty, incorporate auxiliary information, and facilitate interpretability.  For example, in the two-sample test, we compare the log-probability of strings under contexts from its own versus the other group.  We show on synthetic data that the method’s $t$-statistics exhibit the expected null behaviour while maintaining power and, through a replication, that the method is powerful and interpretable. We next introduce text-on-text regression.   Contexts generated around the predictor string are treated as mediating variables between the predictor and outcome strings.  Using negative controls, we then distinguish between semantic and syntactic dimensions of prediction. Analysis of real-world dialogic data illustrates behaviour predicted from a psycholinguistic framework. Theoretically, we provide identification conditions, derive an influence-function decomposition, and show that repeated cross-fitting  of a pivotal statistic yields higher-order efficiency. We derive bounds linking estimation error, context count, and number of cross-fits.  Taken together, context augmentation offers the ability to connect LLMs with longstanding statistical practice.

\end{abstract}
\newpage
\section{Introduction}
\ifarxiv
\else
 \setstretch{2}
\fi
Large language models (LLMs), such as GPT \citep{brown2020language, openai:2023} and BERT \citep{devlin2019bert}, have transformed research across disciplines, including political science, business, law, chemistry, computer science, economics, and beyond \citep[e.g.,][]{Chen2024,noy2023experimental, shah2023creation, argyle:2023}. These models allow working with natural language while preserving syntactic structure and semantic nuance, rather than reducing text to simplified features such as word counts, stems, or pre-defined n-grams \citep{denny2018text}. Despite being probability models,  LLMs remain oddly disconnected from the standard tools of statistical inference:  hypothesis testing, regression modelling, the two-sample test, and so on. Early text methods rely on bag-of-words assumptions or fixed text representations that do not reflect uncertainty \citep{blei2003latent,taddy:13,gentzkow2019text}. More recent approaches attempt inference by treating high-dimensional textual embeddings as fixed inputs to causal estimators or hypothesis tests \citep[e.g.,][]{rodr:spir:2023,veitch2020adapting,wang2021counterfactual}, neglecting the uncertainty in these embeddings themselves. Posterior sampling or bootstrapping could, in principle, address this gap \citep[e.g.,][]{gentzkow2019text}, but these procedures are computationally prohibitive with LLMs, which may involve billions of parameters.

We ask: how can we utilize LLMs as tools of statistical inference, in a manner that connects them with the familiar tools and results of statistical theory? This paper offers a first answer. We propose a method for constructing estimators with known frequentist properties directly from natural language data. The central insight is to treat contexts, which are textual environments in which an observed string might plausibly appear, as latent variables. Using an LLM, we generate these contexts under different conditions, then evaluate how likely the observed string is to appear within each. The method, \emph{context augmentation}, builds on ideas from data augmentation \citep{rubin1976inference,tanner1987calculation}, treating the context around each observed string as a latent variable. We then use an LLM to generate plausible contexts for each observed string, where these contexts serve three roles. First, they reintroduce the uncertainty that traditional embedding methods discard. Second, they function as auxiliary information that allows measuring the differential probability of the observed string across contexts that convey different sorts of information. Third, they allow us to examine the particular sense in which the observed strings operate. This offers a direct connection to the distributional hypothesis in linguistic theory, where similarity in meaning follows from similarity across contexts \citep{harris1954distributional,sahl:2008}.

This may be best explained through example.  For the two-sample problem, we want to know whether a string is in one group or another. To test this, we generate a context from the strings in each group.  We then compare the likelihood that a string will appear in contexts from its own group vs. the other group.  Averaging over the sample yields a $t$-statistic, which we show both analytically and through simulation that it has the expected null distribution.  We avoid self-referential bias, akin to the reflection problem in networks \citep{mans:1993}, by following a repeated cross-fitting strategy \citep{chernozhukov2018double}: strings are evaluated against contexts from a disjoint subsample, the results averaged, and then this process is repeated.

For the regression problem, we consider how the probability of observing a predictor string predicts the probability of observing an outcome string.  We generate contexts around the predictor string, and measure the likelihood of the outcome string appearing outside this context.  These generated contexts mediate the relationship between the predictor and outcome string, allowing us to turn to established frameworks for estimating the effect \citep{pearl2012causal,imai2010general}.  However, rather than intervening on a binary treatment, we construct a more general contrast: moving from informative to uninformative text. Building on ideas from constructed controls \citep{candes2018panning, romano2019deep}, we generate synthetic null variables that preserve structure while removing signal.    We generate syntactically uninformative strings by permuting word order, and semantically uninformative strings by replacing each word with a randomly sampled word of the same part of speech. These negative controls allow us to distinguish the semantic and syntactic dimensions in estimating the impact of a predictor string on an outcome string.

Formally, we provide conditions under which context augmentation identifies target parameters and can generate an estimate with known limiting distribution. Relying on classical frameworks for missing data and potential outcomes \citep{rubin1976inference}, we give precise identification assumptions.   Second, we show that our estimation strategy admits an influence function decomposition and a limiting normal distribution  \citep[e.g][]{vand:1998}. This result isolates three sources of error: sampling variation, variation in generated contexts, and a bias term that arises when the same string is used for both generation and evaluation. This final term motivates our use of repeated cross-fitting.  

We offer two theoretical contributions. First, we demonstrate that when the test statistic is pivotal (e.g., a $t$-statistic), that under some weak regularity conditions, cross-fitting yields a second-order accurate estimator, building on higher-order asymptotic results  and bootstrap theory \citep{hall1992bootstrap, bhattacharya1976normal, swaneepoel2018asymptotic}. Second, we derive a set of bounds that connect how many contexts must be generated, and how many repeated cross-fits, in order to allow for valid inference as a function of the sample size.  To summarize the basic result, repeatedly cross-fitting a pivotal statistic allows us to reduce the order of generated contexts from order $n^{1/2}$ to $n^{1/4}$: a real and practical gain.

We illustrate the method with two applications. First, we test for differences in text responses across experimental treatment groups using data from \citet{egami2022causal}, showing that context augmentation recovers a stronger treatment effect than the original topic model implementation. Second, we introduce a regression framework where both predictors and outcomes are natural language, inspired by psycholinguistic theories of interactive alignment \citep{pickering2004toward,pickering2013integrated}. Using dialogue data, we show that speakers converge to predictable syntax following moments of disagreement, a pattern consistent with integrative repair in conversation.

We turn next to an intuitive discussion of context augmentation.  Section 3 situates context augmentation within relevant literatures. Sections 4 and 5 formally present the setup, identification assumptions, and asymptotic results. Sections 6 and 7 illustrate the method with the two-sample and regression problem, respectively.  A discussion and conclusion follow.

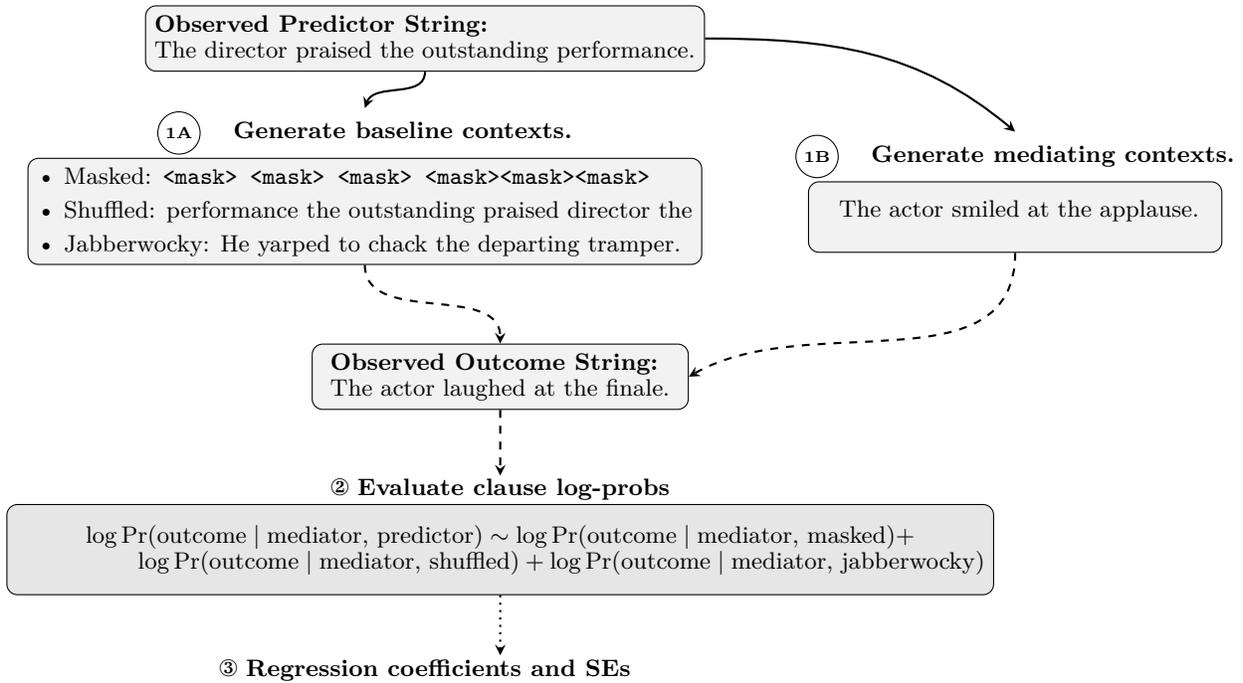
\begin{figure}[p]
  \centering
  \begin{subfigure}{\textwidth}
    \centering
    \begin{tikzpicture}[
  font=\small,
  >=stealth,
  section/.style={font=\bfseries},
  groupbox/.style={draw, rounded corners, fill=gray!5, inner sep=6pt},
  rbox/.style={draw, rectangle, rounded corners, fill=gray!10,
               minimum width=5cm, align=left, font=\footnotesize},
  proc/.style={draw, rectangle, fill=gray!20, rounded corners,
               minimum width=4cm, minimum height=1.2cm, align=center},
  genarrow/.style={->, thick},          
  evalarrow/.style={->, thick, dashed}, 
  aggarrow/.style={->, thick, dotted},  
  stagelab/.style={font=\footnotesize, inner sep=2pt}
]

\node[section,anchor=west] at (-6.3,1) {(a)};
\node[section]            at (0,1) {Two-Sample Test via Context Augmentation};

\node[groupbox] (groups) at (-3.2,-1.7) {
  \begin{tabular}{l}
    \textbf{Group A:}\\
    apple, pear, grape\\[0.3em]
    \textbf{Group B:}\\
    Paris, Berlin, Rome
  \end{tabular}
};

\node (probe) at (0,-1.7) {apple};

\node[rbox] (ctxA) at (3,-0.7) {The tart apple\\tastes great in a pie};
\node[rbox] (ctxB) at (3,-2.7) {apple is the\\capital of France};

\node[stagelab, above =0pt of ctxA] {\ding{172} \footnotesize \textbf{Generate contexts}};

\node[fill=white,inner sep=2pt] (evalA) at (0,-4.5)
 {\footnotesize \textbf{ Evaluate clause log-probs}};
\node[stagelab, left=4pt of evalA] {\ding{173}};

\node[proc] (statA) at (0,-6)
  {Compare across groups\\[0.2em]string-level statistic};
\node (finalA) at (0,-7.5) {\ding{174} \textbf{Sample $t$-statistic}};

\draw[genarrow]  (probe) to[bend left=20]  (ctxA.west);
\draw[genarrow]  (probe) to[bend right=20] (ctxB.west);

\draw[evalarrow] (probe) -- (evalA);
\draw[aggarrow]  (evalA) -- (statA.north);
\draw[aggarrow]  (statA.south) -- (finalA.north);

\end{tikzpicture}
    \label{fig:two-sample}
      \vspace{1em}
  \end{subfigure}
  \vspace{1em}
  \begin{subfigure}{\textwidth}
    \centering
    \begin{tikzpicture}[
  font=\small,
  >=stealth,
  section/.style={font=\bfseries},
  groupbox/.style={draw, rounded corners, fill=gray!5, inner sep=6pt},
  rbox/.style={draw, rectangle, rounded corners, fill=gray!10,
               minimum width=5cm, align=left, font=\footnotesize},
  proc/.style={draw, rectangle, fill=gray!20, rounded corners,
               minimum width=4cm, minimum height=1.2cm, align=center},
  genarrow/.style={->, thick},          
  evalarrow/.style={->, thick, dashed}, 
  aggarrow/.style={->, thick, dotted},  
  stagelab/.style={font=\footnotesize, inner sep=1pt}
]

\node[section,anchor=west] at (-6.3,1) {(b)};
\node[section]             at (0,1) {Text Regression via Context Augmentation};

\node[rbox] (obs) at (-3,-0.5)
  {\textbf{Observed Predictor String:}\\
   The director praised the outstanding performance.};

\node[rbox] (base) at (-3.8,-2.8)
  {• Masked: \texttt{<mask> <mask> <mask> <mask><mask><mask>}\\[0.3em]
   • Shuffled: performance the outstanding praised director the\\[0.3em]
   • Jabberwocky: He yarped to chack the departing tramper.};

\node[rbox, anchor=north east] (med) at (7.6,-2.4)
  {\\$\;\;$
   The actor smiled at the applause.$\;\;$\\};

\node[stagelab, above=0pt of med] (stage1b-label)
  {
    \tikz[baseline=(char.base)]{
      \node[shape=circle, draw, inner sep=1pt, minimum size=16pt, font=\tiny] (char) {\textbf{1B}};
    }
    \quad \footnotesize \textbf{Generate mediating contexts.}
  };

\draw[genarrow] (obs.east) to[out=0, in=140] (stage1b-label.north);

\node[stagelab, above=0pt of base] (stage1a-label)
  {
    \tikz[baseline=(char.base)]{
      \node[shape=circle, draw, inner sep=1pt, minimum size=16pt, font=\tiny] (char) {\textbf{1A}};
    }
    \quad \footnotesize \textbf{Generate baseline contexts.}
  };

\draw[genarrow] (obs.south) to[out=-90,in=70] (stage1a-label.north);

\node[rbox] (out) at (-2,-5)
  {\textbf{Observed Outcome String:}\\
   The actor laughed at the finale.};

\draw[evalarrow] (med.south) to[out=-90,in=30] (out.east);
\draw[evalarrow] (base.south) to[out=-90,in=90] (out.north);

\node[proc] (evalB) at (-2,-7.3)
  {
   $\log\Pr(\text{outcome}\mid\text{mediator},\,\text{predictor}) \sim
   \log\Pr(\text{outcome}\mid\text{mediator},\,\text{masked}) + $\\$\quad\quad\quad\quad\quad\log\Pr(\text{outcome}\mid\text{mediator},\,\text{shuffled}) +\log\Pr(\text{outcome}\mid\text{mediator},\,\text{jabberwocky}) $};

\node[stagelab] (evalBtext) at (-2,-6.5)
  {\ding{173} \textbf{Evaluate clause log-probs}};

\draw[evalarrow] (out.south) -- (evalBtext.north);

\node at (-3,-8.9) {\ding{174} \textbf{Regression coefficients and SEs}};
\draw[aggarrow] (evalB.south) -- ++(0,-0.8);

\end{tikzpicture}
    \label{fig:regression}
  \end{subfigure}
  \caption{\textbf{Intuitive overview of our context-augmentation approach.}}
  \label{fig:intuition}
\end{figure}

\section{Intuition Behind Context Augmentation}

In the two-sample setting, we compare two groups of text, Group A and Group B, as in the top part of Figure \ref{fig:intuition}. An LLM is used to generate multiple contexts around each observed string. For example, given the string ``apple,'' the LLM might generate ``The tart $<<<$STR$>>>$ tastes great in a pie.'' Such a context is clearly aligned with the food domain, resulting in a high probability for “apple,” whereas unrelated strings like “Paris” or “scuba diving” would receive low probabilities. Using these generated contexts, we then construct a statistic for each string that quantifies whether it is more likely to appear under the contexts generated from members of one group versus the other. We avoid a self-referential bias by not comparing strings against contexts generated by that self-same string, and then repeatedly cross-fit.  We show that the $t$-statistics generated from this properly follows the null distribution when Group A and B are the same, but also differentiates between Groups when they are different.  We present pseudocode in Appendix \ref{app:twosamplepseudocode}.

We also apply our framework to the regression setting, illustrated in the bottom part of Figure \ref{fig:intuition}.  Here, we estimate whether observing a predictor string \(s_x\) increases the probability of subsequently observing an outcome string \(s_y\). To capture this effect, we generate contexts around the observed \(s_x\); we then treat these contexts as mediating variables that link the syntactic and semantic features of \(s_x\) to \(s_y\).   To isolate the influence of \(s_x\)’s content, we compare the LLM-assessed probability of \(s_y\) when paired with the informative \(s_x\) against that when \(s_x\) is replaced by a non-informative variant, \(\tilde{s}_x\). We generate non-informative strings in three ways: the masked variant replaces each token in \(s_x\) with a placeholder;  the shuffled variant contains a random permutation of the words; and the jabberwocky variant is constructed by replacing each content word with a gibberish word, preserving syntax but removing semantic meaning. For example, consider the sentence “The director praised the outstanding performance.” Its shuffled variant might be “performance the outstanding praised director the,” while its jabberwocky version could be ``He yarped to catch the departing tramper.''  Holding the generated contexts fixed, we  then regress the log-probability of \(s_y\) given the context and observed predictor on the log-probability of \(s_y\) given the context and the constructed baselines, our negative controls. This approach distinguishes the contributions of semantic content, syntax, and lexical information, even when both predictor and outcome are textual. Appendix \label{app:regressionpseudocode} contains pseudocode.

\section{Literature Review and Motivation}

LLMs, as a new class of models, force us to revisit old questions: how should we quantify uncertainty and conduct inference when the data are textual? How can we use LLMs to construct estimators with verifiable asymptotic guarantees? For numeric data, we have a well-established toolkit to address these questions through estimating equations, robust estimation, and influence function decompositions \citep[see, e.g.,][]{vand:1998,vandervaart1996weak}. Recent semiparametric estimation techniques have successfully managed high-dimensional nuisance functions via regularization, orthogonal moment conditions, and machine learning \citep[e.g.,][]{chernozhukov2014cqiv,Belloni14,chernozhukov2022locally, wage:athe:2017,athe:tibs:2019, farrell2021deep}, building upon classical foundations in profiling, marginalizing, debiasing, and sample-splitting \citep{bick:82,Berger1999,Robins95,murphy2000profile}.

These methods, although theoretically subtle, are well-understood but have not been fully integrated with text data. The key conceptual move in our framework is to treat context as a latent auxiliary variable around each observed string. This connects directly to the classical data augmentation literature \citep{rubin1976inference,tanner1987calculation,dempster1977maximum}, where inference is performed by integrating over unobserved, but simulated,  data points. Here, LLM-generated contexts serve that auxiliary role, allowing us to recover target parameters while also reintroducing uncertainty.

Works utilizing statistical inference with text data turn to two common tools: conformal inference and permutation tests \citep[e.g.,][]{lei2018conformal, PesarinSalmaso2010}. Recent works have applied conformal inference to LLMs \citep[][]{Kumar2023ConformalLLM, Kim2024AdaptiveConformal}. While conformal intervals have many desirable properties, they return valid predictive intervals, whereas we focus instead on inference on population parameters. Permutation tests also offer attractive properties, and have been used in studies involving deep models like LLMs \citep[e.g.,][]{rauba2024distribution, rodr:spir:2023}. However, such tests can be inapplicable or misleading when hypotheses and predictors are correlated \citep[e.g.,][]{Strobl2007}. By augmenting the observed data with contexts that can account for dependencies in the data, our method can avoid this bias.

For the two-sample problem with text, recent methods have often relied on probabilistic topic representations \citep{egami2022causal,roberts2020adjusting}, either through matching or effect estimation across topics. Alternative approaches use kernel-based or neural methodologies, such as MMD and deep-learning-based two-sample tests \citep{gretton2012kernel,schrab2023,kirchler2020two}, or entropy-based model evaluations \citep{xu2025two}. However, these approaches still largely rely on fixed representations. Context augmentation moves beyond these fixed embeddings by incorporating linguistic variability through generated contexts,  reintroducing uncertainty and  contextual information.

In regression settings, existing methodologies predominantly handle text either as an outcome \citep[e.g.,][]{rodr:spir:2023,roberts2014structural} or as a predictor \citep[e.g.,][]{debartolomeis2025efficient, farrell2021deep, taddy:13,blei2007supervised}, without supporting text-to-text inference explicitly. Methods such as Structural Topic Models, supervised LDA \citep{blei2007supervised}, multinomial inverse regression \citep{taddy:13}, and embedding-based regression frameworks focus on numeric or categorical outcomes. No existing method, to our knowledge,  supports text-to-text regression with valid inference. Additionally, we introduce LLM-generated negative controls, the "jabberwocky" and shuffled variants, to distinguish semantic from syntactic contributions.

We recover second-order efficient estimates through repeated cross-fitting of a pivotal statistic.  Unlike 
studies that rely on plug-in corrections of higher-order bias terms in U-statistics \citep[e.g.,][]{vand:2014,Robins2008,li:tche:vand:2011}, we leverage pivotality to directly eliminate the leading bias term in an Edgeworth expansion  \citep[akin to][]{hall1992bootstrap}. The closest antecedent to our approach is \citet{swaneepoel2018asymptotic}, who eliminate leading bias terms using subsampled plug-in estimates in a Cornish-Fisher expansion. Our contribution extends this result into a more general, semiparametric, textual setting.

\section{Setup and Notation}

We formalize the problem using notation tailored to text-valued data, distinguishing between observed strings, the contexts in which they are embedded, and the clause functions linking the two. Our framework aggregates over contexts within each string and then across strings to the sample level, in a manner compatible with standard tools of statistical analysis. Table~\ref{tab:notation} provides a summary before the formal notation is introduced.
\begin{table}[tbp!]
\small
\centering
\begin{tabular}{ll}
\hline
\multicolumn{2}{l}{\textbf{Dimensions}}\\[3pt]
$d$ & components per string vector \\
$p$ & covariates per string \\
$q$ & conditioning events, one context per event $(k=1,\dots,q)$\\
\hline
\multicolumn{2}{l}{\textbf{Variables}} \\[3pt]
$s,\;s_i$ & $s$: $d$-vector of random strings; $s_i$: $i$th observed string ($d$-vector) \\[3pt]
$\mathbf x_i$ & Covariate vector of length $p$ for $s_i$ \\[3pt]
$n,\;n_c$ & Number of strings; contexts per string $(j=1,\dots,n_c)$ \\[3pt]
$\mathcal E,\;\mathcal E^{(k)}$ & Set of events; $k$-th event, $k=1,\dots,q$ \\[6pt]
\multicolumn{2}{l}{\textbf{Latent Contexts}} \\[3pt]
$c,\;c_{j;i},\;c^{\mathcal E}$ & $c_{j;i}$: $j$-th context for $s_i$;
$c^{\mathcal E}$: context vector under event $\mathcal E$ \\[6pt]
\multicolumn{2}{l}{\textbf{Parameters and Models}} \\[3pt]
$\theta$ & Target estimand (scalar or vector) \\[3pt]
$\widehat{\mathcal M},\;\widehat{\mathcal M}^{c}$ & Scoring model; context-generation model \\[6pt]
\multicolumn{2}{l}{\textbf{Operators \& Functions}} \\[3pt]
$Cl(s,c,\mathcal E,\widehat{\mathcal M})$ & Clause function: returns a $d\times q$ score matrix \\[3pt]
$Str(s,\mathcal E,\widehat{\mathcal M})$ & Aggregates $Cl$ over contexts (operator $\mathcal A$) \\[3pt]
$\mathcal A(\cdot),\,T(\cdot)$ & $\mathcal A$: within-string aggregation; $T$: sample-level aggregation to $\theta$ \\
\hline
\end{tabular}
\caption{\textbf{Key notation.}  Unless specified, vectors and scalars are both in italics; subscripts index observations or components.}
\label{tab:notation}
\end{table}


\subsection{Data and Variables}

We assume that the researcher observes $n$ realizations of a $d$-dimensional vector-valued random variable \(\mathbf{s}\), denoted \(\{\mathbf{s}_i\}_{i=1}^n\), with $\mathbf s_i = (s_{i1}, s_{i2}, \ldots, s_{id})^\top$.  This, along with a vector of $p$ string-level covariates for each observation, $\mathbf x_i$, comprises the observed data. 

We will measure how the string-probability changes across a set of conditioning events $\mathcal E = \{\mathcal E^{(1)}, \ldots, \mathcal E^{(q)}\}$.  In the two sample problem, the event is sample membership; in the regression problem, it is presence of a predictor string.  While we cannot evaluate the string-probability conditional on an event, we will use the large language model (LLM) $\wmM$ to generate \emph{event-contexts} that capture the information in the event:
\begin{equation}
\mathbf c^\mathcal E= \mathbf c\mid \mathcal E, \wmM.
\end{equation}
where $\mathbf c$ is a $q$-dimensional text-valued vector.  Each element of the context vector corresponds to an element of the event set,  such that the clause function evaluates the string across each joint context/event pair. We generate $n_c$ contexts per string, with context $j$ generated off string $\mathbf s_i$ as $\mathbf c_{j;i}$. These event-contexts will provide the auxiliary information connecting strings to events.



For the regression problem, we generate contexts off one set of strings, the predictor strings then evaluate them against another, the outcome strings.  In the two-sample problem, the event--sample-membership--is operationalized by the strings in each event.  A full-sample analysis, in which strings are evaluated against contexts generated by those same strings, induces a self-referential bias \citep[analogous to the ``reflection problem'' of][]{mans:1993}.  We address this bias in two stages: cross-fitting eliminates the lead bias term and pivotality the second.  First, though, we turn to constructing our estimand and estimate.

\subsection{The Estimand}

We construct our estimand, $\theta$, in two steps.  First, we aggregate over the context distribution to  recover a string-level object, and then over the string distribution to obtain a population-level parameter.  It will be useful at times to write this parameter as a functional of the joint distribution of strings and contexts by event, i.e. $\theta\doteq \theta(F_{\mathbf s, \mathbf c ; \mathcal E})$.

Our basic object is the \textit{clause function}, which evaluates each of the $d$ components of the string vector against each element of the $q$-dimensional context vector, and then maps each string-context pair to a real number,  returning a $d \times q$ matrix of scores:
\[
Cl(\mathbf s, \mathbf c, \mathcal E, \wmM):  \mathcal S^d \times \mathcal C^q \mapsto \Re^{d \times q}
\]
Evaluation will be done by a scoring model, $\wmM$.  For simplicity, a single model for generation and scoring is used throughout this paper.   The model, $\wmM$, includes the tokenizer, weights, and other components required for text evaluation. For generation, additional hyperparameters, such as temperature and prompt formatting, are also included. Inference will be dependent on these values, which we make explicit by conditioning on $\wmM$ at each step.  While we focus on log-probability scores, clause-level evaluations can also be passed to a secondary model that returns summary measures, such as sentiment or ideology. Appendices \ref{app:twosamplepseudocode}-\ref{app:regressionpseudocode} contain pseudocode showing our prompt construction and algorithm flow.

In constructing $\theta$, we first apply the clause function and aggregate over contexts with a user-specified aggregation operator $\mathcal A$,
\[
Str(\mathbf{s}, \mathcal{E}, \wmM) = \mathcal{A}\circ Cl(\mathbf{s}, \mathbf{c}, \mathcal{E}, \wmM),
\]
returning a string-level functional.  Given the high-dimensional, and possibly erratic, nature of LLMs, this aggregation encompasses means but also robust alternatives, like medians, quantiles, ranks, trimmed means, or other robust location statistics.  We then map this string-level object to our target parameter through the operator $T$, also user-specified,
\[
\theta = T\circ Str(\mathbf{s}, \mathcal{E}, \wmM)
\]
The $T(\cdot)$ functional is also flexible, allowing for the same robust alternatives as $\mathcal A$ and smooth transformations thereof, such as logarithms.  Formally, we will allow any transformation that admits a first-order Hadamard derivative, allowing for an asymptotic linearization.

\subsection{The Estimate}

Plug-in estimates are denoted with a hat. For observation \(\mathbf{s}_i\) and its associated contexts \(\{\mathbf{c}_{j;i}\}_{j=1}^{n_c}\), we construct the estimate
\begin{align}
\widehat{Str}(\mathbf{s}_i, \mathcal{E}, \wmM) &= \mathcal{A}\circ \{Cl(\mathbf{s}_i, \mathbf c_{j;i}, \mathcal{E}, \wmM)\}_{j=1}^{n_c}\\ &\doteq \widehat{Str}_i^{\mathcal E}.
\end{align}
where we adopt the compact notation in the second line when there is no ambiguity.
The estimator for \(\theta\) is then
\[
\widehat \theta = T\; \circ\; \{ \widehat{Str}_i^{\mathcal E}\}_{i=1}^n.
\]
The variance in $\widehat \theta$ enters from two sources: the first within-string variance driven by contexts and the second cross-string variance over the sample.

\section{Identification and Estimation}
We next turn to four sets of results on the identification and estimation of our target parameter, $\theta$.  We first state a set of identification assumptions analogous to those from the literatures on missing data and causal inference \citep{rubin1976inference}.  Second, we derive the limiting distribution of our estimator in a way that characterizes bias induced through self-referential event-context generation.  Third, we follow standard arguments in semiparametric theory and show that a subsampling strategy eliminates bias to first order \citep[see, e.g.][]{chernozhukov2018double, vand:1998}. Finally, we provide two extensions. In the first, we extend results on higher-order efficiency by showing that cross-fitting a pivotal statistic can eliminate a second-order bias term—a new result  in our setting \citep[e.g.,][]{vand:2014, Robins2008, li:tche:vand:2011, swaneepoel2018asymptotic}.  In the second, we provide a bound balancing the rate requirement on the nuisance against the number of sampled contexts.  We show how the rate can be met through a combination of computationally-expensive generated contexts and computationally inexpensive  repeated cross-fits.

\subsection{Identification}

We connect strings to events using $\mathbf c^{\mathcal E}$ as auxiliary information. Under the following assumptions, differences in $\theta$ across separate sets of event-contexts imply differences in the underlying string-probabilities:
\begin{enumerate}
    \item \textbf{Overlap:}  
    For events $\mathcal E, \mathcal E^\prime$ and all $s$ in the support of $F_{s|\mathcal E} \cup F_{s|\mathcal E'}$,
    \[
      \operatorname{supp}(F_{c|s, \mathcal E}) = \operatorname{supp}(F_{c|s, \mathcal E'}).
    \]

    \item \textbf{Weak Ignorability:} For any two event sets \( \mathcal{E} \) and \( \mathcal{E}^\prime \),
    \[
    Cl(\mathbf{s}, \mathbf{c}^{\mathcal{E}}, \mathcal{E}, \wmM) 
    =
    Cl(\mathbf{s}, \mathbf{c}^{\mathcal{E}}, \mathcal{E}^\prime, \wmM).
    \]

    \item \textbf{Injectivity:} The mapping \( \theta(\cdot) = T \circ \mathcal{A} \) is injective.
\end{enumerate}

\paragraph{Lemma (Identification).}  
\emph{Under these assumptions, 
\[
\theta(F_{\mathbf{s},\mathbf{c}^{\mathcal{E}}};\mathcal{E}) \ne \theta(F_{\mathbf{s},\mathbf{c}^{\mathcal{E}'}};\mathcal{E}') 
\quad\Rightarrow\quad 
F_{\mathbf{s}|\mathcal{E}} \ne F_{\mathbf{s}|\mathcal{E}'}.
\]
}

\begin{proof} See Appendix \ref{proof:identification}.
\end{proof}

Our identification assumptions adapt classic strategies utilizing latent variables and auxiliary information to LLM-generated contexts \citep{tanner1987calculation, rubin1976inference, dempster1977maximum}.   The Overlap condition is mild in this setting, since modern LLMs can generate virtually any string in response to a prompt:  the support of contexts will cover the support of observed strings. This guarantees that differences in string distributions are due to changes attributable to different events, and not lack of overlap across events.  Weak Ignorability only requires ignorability of the event given $c^\mathcal E$ when evaluated in the clause function, as opposed to the conditional independence in Strong Ignorability  \citep[see, e.g.][Technical Comment 2.1]{HernanRobins2025}.  We turn now to estimation and inference.

\subsection{Estimation and Asymptotics}

We present three main asymptotic results. First, we set up $\theta$ as the solution to an estimating equation and then generate a decomposition that isolates its sources of variance.  Second, we discuss two standard ways to guarantee convergence and asymptotic unbiasedness: by a Donsker assumption or through a split-sample approach.  Third, we show that, if the target statistic is pivotal, repeated cross-fitting generates a higher-order efficiency.  \citet[][Corollary 3.3 and subsequent text]{chernozhukov2018double} analyze a fixed number of cross‑fits; we show that one can let the number of repeated cross-fits grow with $n$ and still control the error.  We  then balance this against the number of contexts per string $n_c$, needed to achieve a target estimation error rate.    Rather than using plug-in corrections of higher order bias terms in U-statistics \citep[e.g.,][]{vand:2014, Robins2008, li:tche:vand:2011}, we leverage pivotality to eliminate the leading term in an Edgeworth expansion—similar in spirit to \citet{hall1992bootstrap}. Our work is closest to  \citet{swaneepoel2018asymptotic}, who eliminate the lead bias term by using subsampled plug-in estimates in a Cornish-Fisher expansion. Our contribution extends this result to a  general, semiparametric setting with application to text-valued data.

\subsection{Setup and Influence Function Decomposition}

We take as the target parameter, $\theta$, defined above, and the nuisance the distribution of event-conditioned contexts, $\eta = F_{c^{\mathcal E}}$ and $\widehat \eta = \widehat F_{c^{\mathcal E}}$. The complete data consist of observed strings and generated contexts, $\{Z_i\}_{i=1}^n = \{\mathbf{s}_i, \{\mathbf{c}_{j;i}^{\mathcal E}\}_{j=1}^{n_c}\}_{i=1}^n$, with $\widehat F_{c^{\mathcal E};i}$ the estimated distribution of  $\mathbf{c}_{j;i}^{\mathcal E}$.   
The following assumptions allow us to write $\theta$ as the solution to an estimating equation and then develop a von Mises expansion around it:

\begin{assumption}\label{assn:zest}
\begin{enumerate}
\item \textbf{Sampling:}  The string-level statistics $\widehat{\text{Str}}_i^{\mathcal{E}}$ are independent across observations, and the contexts used to construct each $\widehat{\text{Str}}_i^{\mathcal{E}}$ are independent conditional on the string $\mathbf{s}_i$.
\vspace{2pt}
\item \textbf{Population Solution:}  The parameter $\theta \in \Re^d$ solves 
\[
\Phi(\theta, \eta=\eta_0) = \mathbb{E}[\phi_{\theta, \eta=\eta_0}( Str_i^{\mathcal E})] = 0.  
\] 
with  Jacobian $\partial_\theta \Phi(\theta, \eta)$ invertible uniformly in a neighborhood of $\theta_0$.
\vspace{2pt}
\item \textbf{First-Order Expansion:}  The map $\phi_{\theta, \eta}$ is Hadamard differentiable in $\eta$ and pathwise (Gateaux) differentiable in $\theta$, both tangentially to  a subset of $L_2(P)$  that contains the closure of the image of  $\mathcal A\dot Cl$.
\vspace{2pt}
\item \textbf{Differentiable in Quadratic Mean:}  The model is differentiable in quadratic mean  in $\theta$, uniformly in a neighborhood of $(\theta_0, \eta_0)$.
\vspace{2pt}
\item \textbf{Stochastic Equicontinuity:}  The  process $\widehat{\Phi}(\theta, \eta) = \frac{1}{n} \sum_{i=1}^n \phi_{\theta, \eta}(\widehat{Str}_{i; n_c}^{\mathcal E})$ is stochastically equicontinuous uniformly in a neighborhood of $\theta_0$. 
\end{enumerate}
\end{assumption}
These mirror standard assumptions in Z-estimation see \citet[e.g.][Thm. 25.57]{vand:1998} or \citet[e.g.][Sec 2.1]{chernozhukov2018double}, giving us a decomposition into three sources of error: sampling error, context-level error, and self-referential bias.
\begin{result}

Our estimate decomposes into the following form, 
\begin{align}
\sqrt{n}(\widehat{\theta} - \theta_0) 
&= \sqrt{n} \bigg\{ 
\underbrace{(\widehat{\theta}_{\text{Str}, \eta_0} - \theta_0)}_{\text{Sampling variation}} 
+ \underbrace{(\widehat{\theta}_{\widehat{\text{Str}}, \eta_0} - \widehat{\theta}_{\text{Str}, \eta_0})}_{\text{Context-level error}}
+ \underbrace{(\widehat{\theta}_{\widehat{\text{Str}}, \widehat{\eta}} - \widehat{\theta}_{\widehat{\text{Str}}, \eta_0})}_{\text{Self-referential bias}} 
\bigg\} \notag \\
\end{align}
Which, under  Assumption \ref{assn:zest}, admits the representation
\begin{align}
&= \left( \partial_\theta \Phi(\theta_0, \eta_0) \right)^{-1} \cdot \frac{1}{\sqrt{n}} \sum_{i=1}^n \bigg\{ 
\underbrace{\phi_{\theta_0, \eta_0}(\text{Str}_i^{\mathcal{E}})}_{\text{Sampling variation}} \notag \\
&\quad\quad\quad + \underbrace{\partial_{\text{Str}} \phi_{\theta_0, \eta_0}(\text{Str}_i^{\mathcal{E}}) \left[ \widehat{\text{Str}}_{i; n_c}^{\mathcal{E}} - \text{Str}_i^{\mathcal{E}} \right]}_{\text{Context-level error}} \notag \\
&\quad\quad\quad\quad + \underbrace{\partial_{\eta} \phi_{\theta_0, \eta_0}(\widehat{\text{Str}}_{i; n_c}^{\mathcal{E}}) \cdot \left\| \widehat{F}_{c^\mathcal{E}; i} - F_{c^\mathcal{E}; i} \right\|_\infty}_{\text{Self-referential bias}} 
\bigg\} + o_p(1),
\end{align}
which, when useful, we will write the aggregated influence function as
\begin{align}
&\doteq  \frac{1}{\sqrt n} \sum_{i=1}^n \psi_{\theta, \eta}(Z_i) + o_p(1).
\end{align}

\paragraph{Proof.} See Appendix \ref{app:vonmises}.  

\end{result}

This decomposition helps us distinguish between different directions in which the estimate may fluctuate around the population minimizer: sample-variation, context-variation, and estimation error on the event-context distribution.  The first two influence functions can be handled through standard parametric arguments. The third one involves a nonparametric element, which may induce a persistent bias term.

 The following assumption will guarantee a well-behaved limiting distribution:
\begin{assumption}\label{assn:control}
\begin{enumerate}
\item  \textbf{Lyapunov Tail Condition:} The influence function $\psi_{\theta, \eta}$ has $2+\epsilon$ finite moments for some $\epsilon>0$.
\item  \textbf{Pointwise Convergence:}  The string  and error estimate converge pointwise as $\widehat Str_i^\mathcal E \plim Str_i^\mathcal E$ and, at each context $\mathbf c$,  $\widehat{F}_{c^\mathcal{E}}(\mathbf c) \plim F_{c^\mathcal{E}}(\mathbf c)$.
\item \textbf{Uniform bound:} The error on the event context distribution vanishes uniformly over event-contexts and strings as
\[
\frac{1}{\sqrt{n}} \sum_{i=1}^n \partial_{\eta} \phi_{\theta_0, \eta_0}(\widehat{\text{Str}}_{i; n_c}^{\mathcal{E}}) \cdot \left\| \widehat{F}_{c^\mathcal{E};i} - F_{c^\mathcal{E};i} \right\|_\infty \xrightarrow{p} 0
\]

\end{enumerate}
\end{assumption}

\begin{result}[Limit Theorem]\label{thm:clt}
Under Assumptions~\ref{assn:zest} and~\ref{assn:control},  
\[
\sqrt{n}(\widehat{\theta} - \theta_0) \rightsquigarrow \mathcal{N}(0, \Omega),
\]
where the asymptotic variance is given by
\[
\Omega = \left( \partial_\theta \Phi(\theta_0, \eta_0) \right)^{-1} \cdot \mathbb{E} \left[ \phi_{\theta_0, \eta_0}(\text{Str}_i^{\mathcal{E}}) \phi_{\theta_0, \eta_0}(\text{Str}_i^{\mathcal{E}})^\top \right] \cdot \left( \partial_\theta \Phi(\theta_0, \eta_0) \right)^{-1\top}.
\]

\paragraph{Proof.} See Appendix \ref{app:CLT}.  
\end{result}

Controlling the first two influence functions is a standard parametric problem.  The third term requires a bit more care, as the nonparametric element may not achieve the parametric rate.  Several strategies here include a classical no-bias condition for asymptotic normality \citep[][Condition 25.52]{vand:1998}, constraining the complexity of the function to be in a Donsker class.  

In many common machine learning cases, this Donsker assumption is provably invalid.  Recent years have found renewed interest in subsampling methods \citep{bick:82, vand:1998, politis1999subsampling, chernozhukov2018double}, where disjoint sets of data are used to estimate the nuisance and conduct inference.  Under this strategy, the rate requirement on the estimation error on the nuisance term can be reduced from $o_p(n^{-1/2})$ to $o_p(n^{-1/4})$, such that the Assumptions \ref{assn:control} can hold under
\[
\frac{1}{n} \sum_{i=1}^n \left\| \widehat{F}_{c^\mathcal{E};i} - F_{c^\mathcal{E};i} \right\|_\infty = o_p(n^{-1/4}).
\]

Key to achieving this rate is using different subsets of the data to estimate $\widehat \theta=\widehat F_{c^{\mathcal E}}$  and conduct inference on $\theta$.  In the regression problem, we do this naturally: contexts are generated off of predictor strings, while inference is done using a distinct set of outcome strings.  In the two-sample problem, where strings are evaluated against contexts by strings in each group, we adjust for self-referential bias using through subsampling: we split the data in equal-sized subsamples, $\mathcal I^{1}, \mathcal I^{2}$.  We use $\mathcal I^1$ for generating contexts, $\mathcal I^2$ for inference on $\theta$, and then cross-fit, where we swap roles and average.  We then repeatedly cross-fit, taking the mean or median over cross-fits in order to average over the peculiarities of a given split, and then take a mean or median over repeated cross-fits to recover location and uncertainty estimates \citep{chernozhukov2018double}.   The  cross-fitting approach is motivated by a desire to reduce the rate-requirements on nonparametric nuisance terms.  We turn next to a further refinement.

\subsection{Higher-Order Accuracy and Computational Efficiency via Pivotal Statistics}

The generation of contexts and evaluation of clauses is computationally intensive, while the downstream steps--regressions or averaging over contexts--are much less costly.  To connect these tradeoffs with the sample size, we provide a result with two components. First, following the logic of \citet{hall1992bootstrap}, repeated cross-fitting of a pivotal statistic leads to a higher-order efficient estimate. In practice, this reduces the convergence rate required on our nuisance term \(n^{-1/4}\) to \(n^{-1/8}\).  We then bound the lead bias,  connecting the (expensive) number of generated contexts, \(n_c\), and the inexpensive number of repeated cross-fits or bootstraps \(M=R\) or \(B\) with the requirement on the approximation error.  This repeated cross-fitting or bootstrapping strategy can be used to reduce the number of contexts needed to guarantee the convergence rate necessary for valid inference.

\begin{assumption}[Pivotal Limit]
Beyond Assumptions 1–3, suppose:
\begin{enumerate}
  \item \textbf{(Pivot)} There exists a rate \(r_n\to\infty\) and a known distribution \(G\), such that
  \[
    r_n\bigl(\widehat\theta_n - \theta_0\bigr)\;\overset{d}{\longrightarrow}\;G,
  \]
  and \(G\) involves no unknown parameters.

  \item \textbf{Cramér’s Condition on \(G\):}  The characteristic function of \(G\) is non‑zero in a neighborhood of the origin.
 \item \textbf{Tail Condition:}  The influence function $\psi_{\theta, \eta}$ has $4+\epsilon$ finite moments for some $\epsilon>0$.
\end{enumerate}
\end{assumption}
\begin{result}[Convergence Rates under Pivotality]\label{result:pivot}
Under Assumptions 1–3 and the Pivotal Limit condition, suppose we estimate \(\theta\) by either \(R\) repeated cross‑fits or \(B\) bootstrap replicates, denoted as $M$, drawing \(n_c\) contexts per string.  Then:

\begin{enumerate}
  \item \textbf{Second‑order convergence via pivot.}  
    Pivotality reduces the nuisance convergence rate from \(n^{-1/4}\) to \(n^{-1/8}\).
  \item \textbf{Uniform bias bound.}  
    The lead bias term can be controlled uniformly as
    \[
      \mathrm{Bias}
      =
      O_p\!\Bigl(\sqrt{\tfrac{\ln n_c\;\ln\!\ln M}{\,n_c\,M\,}}\Bigr),
    \]
    where \(M=R\) (cross‑fits) or \(M=B\) (bootstrap).  Our analysis is agnostic over these, as both act as symmetric U-statistics over the data.\footnote{For an earlier connection of the behaviour of  $n/2$-subsampling and the bootstrap, see \citet[][p.1178]{shao1989general}.}
    
    Therefore, driving the bias below \(n^{-\gamma}\) (with \(\gamma=\tfrac14\) in the standard semiparametric setting or \(\gamma=\tfrac18\) under pivotality), requires
    \[
      (n_c\,M)^{1+\epsilon}\;\gg\;n^{2\gamma}
      \quad\text{for some small }\epsilon>0.
    \]
\end{enumerate}

\noindent\textbf{Proof.} See Appendix \ref{app:pivotresult}.
\end{result}

\paragraph{Interpretation and Computational Guidance.}

These results supply a unified theoretical rationale for balancing the computational burden of costly context generation against virtually free inference repetitions. Once contexts are drawn and scored, increasing the number of cross‑fits \(R\) or bootstrap replicates \(B\) incurs negligible overhead yet yields equivalent bias control. Under pivotality, the required convergence rate halves—from \(n^{-1/4}\) to \(n^{-1/8}\)—so that a moderate \(n_c\) suffices when paired with a sufficiently large \(R\) or \(B\).

Crucially, these conclusions hold for any pivot, including a normal limit but also the $F$ or chi-square, making the approach broadly applicable in text settings where effect sizes are on a transformed scale. In practical terms,  inexpensive inference repetitions can improve the accuracy and help achieve second‑order accuracy under minimal computational expense.


\section{The Two-Sample Problem via Context Augmentation}
\label{sec:two_sample_problem}

We apply context augmentation to test whether two groups of string-valued data differ in ways  captured by an LLM.  We denote as $c^G$ clauses that could contain strings in group $G \in \{A, B\}$.  The clause function is then
\[
Cl({s}, {c}, \mathcal{E}=G,\wmM) = \log \Pr( s| c^G),
\]
the log probability of observing string $s$ in context $c^G$, as scored by the model.  We then construct the string function as a two-vector, with elements given by the mean log probability in each group,
\[
Str(s, \mathcal{E}=\{A,B\}, \wmM) = 
\begin{bmatrix}
\mathbb{E}_{\mathbf{c}^A} [ \log \Pr( s| c^A)] \\
\mathbb{E}_{\mathbf{c}^B} [ \log \Pr( s| c^B)]
\end{bmatrix}.
\]

Define 
\[
\theta^A = \E(\mathbb{E}_{\mathbf{c}^A} [ \log \Pr( s| c^A)] - \mathbb{E}_{\mathbf{c}^B} [ \log \Pr( s| c^B)] | s \in A)
\]
and $\theta^B$ similarly for group $B$.  Our parameter is then
\[
\theta = \theta^A - \theta^B
\]
and we test the null hypothesis $\mathcal H_0: \theta=0$.

\paragraph{Estimation.}   We estimate the test statistic using a repeated cross-fitting procedure designed to eliminate self-referential bias. First, we randomly partition the data into two folds: one fold for generating contexts and the other for conducting inference. For each string in the inference fold, we evaluate its probability  under the contexts in group $A$ and group $B$ generated from strings in the other fold.  This process is repeated in both directions (i.e., swapping the roles of the two folds), and the resulting $t$-statistics are averaged and rescaled to correct for the sample-splitting. To stabilize results and smooth out any randomness from the fold assignment, we repeat this procedure and average over 25 repeated cross-fits. For details, see Appendix \ref{app:estimation}.
\subsection{Empirical Demonstration}
\label{subsec:two_sample_empirical}

We evaluate our approach using synthetic data generated by GPT-4 across five categories (animals, body parts, cities, food, plants). For each category, we sample 100 multi-word strings and generate 10 contexts per string using FLAN-T5-XXL.\footnote{WordNet over-represents Western countries; to broaden coverage we prompted GPT-4 to produce groups with broader geographic coverage.}

Contexts are generated using a procedure summarized in Table~\ref{tab:aligned_pipeline_with_output}. Each stage builds on the previous, allowing us to build up contexts.   The definition prompt enforces some similarity between the left and right contexts.  Constructing, and storing, left- and right-contexts separately makes evaluation easier, making it simple to loop strings between the contexts on each side.

\begin{table}[tb!]
\singlespacing
\centering
\small
\resizebox{.9\textwidth}{!}{%
\begin{tabular}{@{}c l p{6cm} p{6cm}@{}}
\toprule
\textbf{Opt.} & \textbf{Stage}      & \textbf{Prompt Snippet}                                                                      & \textbf{Example Output}                                                           \\
\midrule
\multirow{4}{*}{A}
  & Definition     & \texttt{Provide a clear and concise summary or definition of 'leopard'.}                                    & in the animal kingdom        \\[0.5em]
  & Left Context   & \texttt{Using the definition '\{sense\}', provide a sentence fragment before the word 'leopard'.}             & in the animal kingdom is a genus of four legged animals .           \\[0.5em]
  & Right Context  & \texttt{Using the definition '\{sense\}', provide a sentence fragment after '\{left\} leopard'.}               & is a genus of mammals that contains those very big, spotted cats.                                 \\[0.5em]
  & Output         &                                                                                                           &  in the animal kingdom is a genus of four legged animals .  <<<STR>>> leopard in the animal kingdom \\
\midrule
\multirow{4}{*}{B}
  & Definition     & \texttt{Provide a clear and concise summary or definition of 'leopard'.}                                    & a species that lives in warm temperate regions                                                                     \\[0.5em]
  & Left Context   & \texttt{Using the definition '\{sense\}', provide a sentence fragment before the word 'leopard'.}             & A species is a living thing that lives in a particular environment.                                                          \\[0.5em]
  & Right Context  & \texttt{Using the definition '\{sense\}', provide a sentence fragment after '\{left\} leopard'.}               & Leopards are a species that live in warm temperate regions.                                            \\[0.5em]
  & Output         &                                                                                                           & A species is a living thing that lives in a particular environment. <<<STR>>> Leopards are a species that live in warm temperate regions.                                                         \\
\midrule
\multirow{4}{*}{C}
  & Definition     & \texttt{Provide a clear and concise summary or definition of 'leopard'.}                                    & The person or animal who steals.                                                                   \\[0.5em]
  & Left Context   & \texttt{Using the definition '\{sense\}', provide a sentence fragment before the word 'leopard'.}             & The thief , a leopard, slithered out of the room quickly.                                                                               \\[0.5em]
  & Right Context  & \texttt{Using the definition '\{sense\}', provide a sentence fragment after '\{left\} leopard'.}               & thief                                                                     \\[0.5em]
  & Output         &                                                                                                           &The thief , a leopard, slithered out of the room quickly. <<<STR>>> thief                                              \\
\bottomrule
\end{tabular}}
\caption{\textbf{Context generation.} Examples of contexts generated around the word \texttt{leopard}.  The first example is biological, the second deals with its environment, and the third uses leopard as a  metaphor. The appearance of the target word in the context illustrates the need for cross-fitting.  }
\label{tab:aligned_pipeline_with_output}
\end{table}

\begin{figure}[t!]
    \centering
    \includegraphics[width=0.6\textwidth]{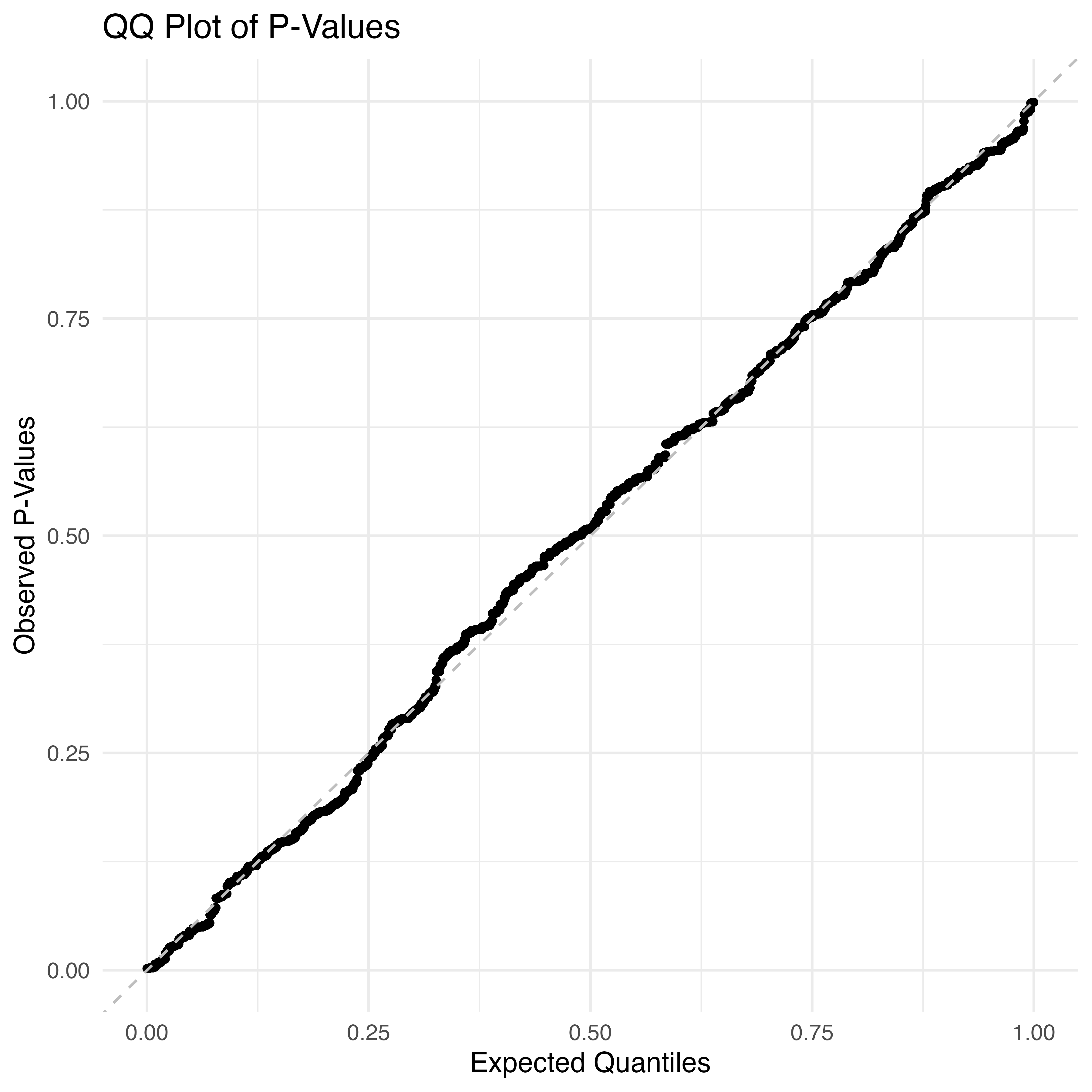}
    \caption{\textbf{Null calibration.} QQ-plot of $p$-values from within-category comparisons.  Alignment with the diagonal illustrates valid null calibration.}
    \label{fig:qqplot}
\end{figure}

\paragraph{Results.}  
We assess our method in two ways. First, for null calibration, strings from the  semantic category are randomly split into two groups.  Figure~\ref{fig:qqplot} confirms that under the null, the p-values from our test are uniformly distributed (the QQ-plot aligns with the diagonal), indicating correct calibration.

\begin{figure}[t!]
    \centering
    \includegraphics[width=0.75\textwidth]{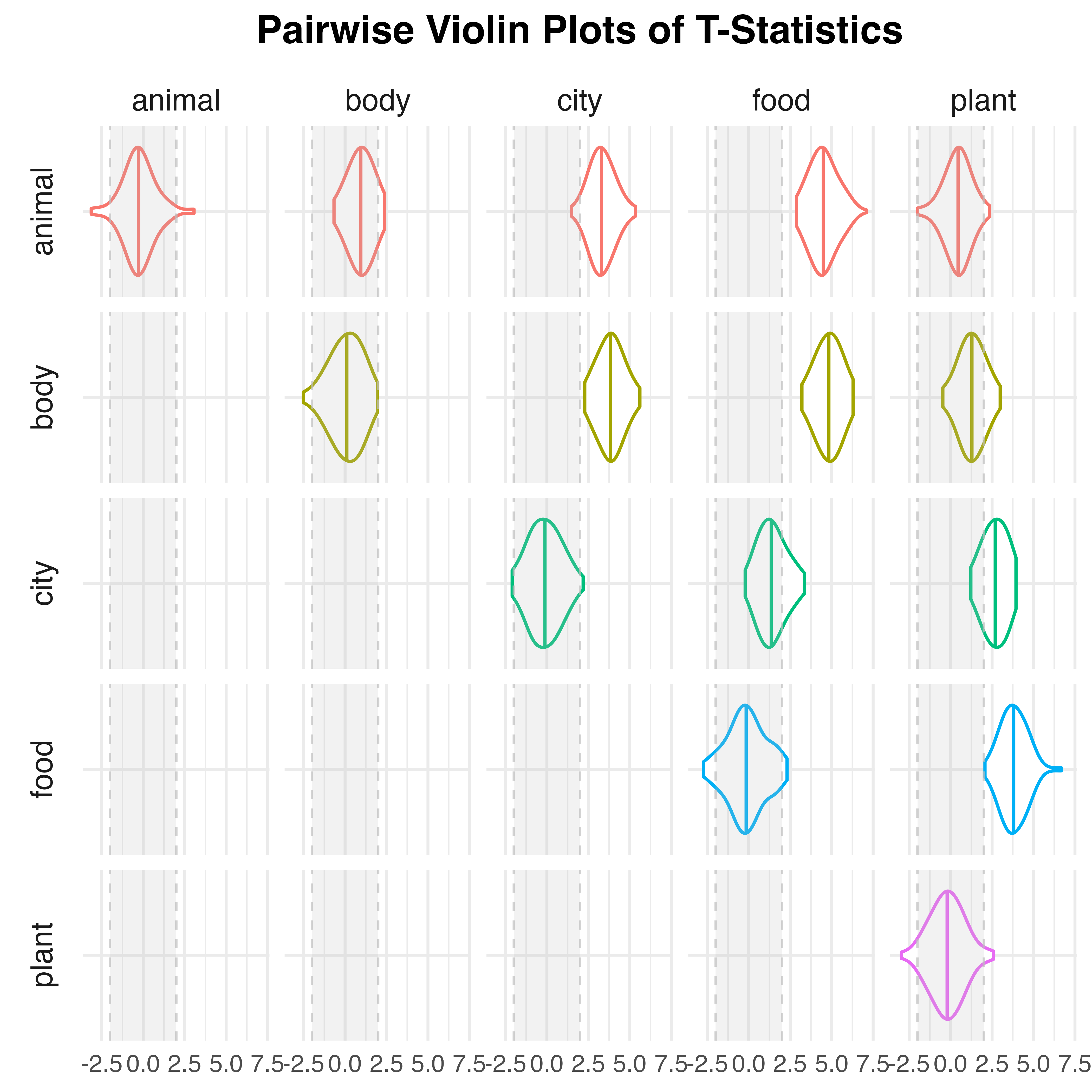}
    \caption{\textbf{Power.} Violin plots of pairwise $t$-statistics for cross-category comparisons. Shaded regions represent non-significant $t$-statistics ($|t| < 2$). The distributions highlight clear differentiation among semantic categories.}
    \label{fig:violinplot}
\end{figure}

Figure~\ref{fig:violinplot} illustrates the methods power, displaying pairwise comparisons across different categories.  The results show strong and systematic differences, with most test statistics exceeding conventional thresholds ($|t| > 2$), demonstrating that the method is sensitive to meaningful semantic distinctions. Analytic variances closely matched bootstrap estimates, supporting the robustness of the inference.

\section{Regression via Context Mediation}
\label{sec:regression_context_mediation}

We consider a regression setting where both the predictor \(s_x\) and outcome \(s_y\) are text. Our goal is to measure whether observing \(s_x\) increases the probability of subsequently observing \(s_y\). The effect of an input string \(s_x\) on an outcome string \(s_y\) is assumed to occur through latent contexts, which we generate using a large language model (LLM). These contexts, derived from the informative \(s_x\), embed both the semantic and syntactic features of the text in a high-dimensional space. To assess how the characteristics of \(s_x\) influence \(s_y\), we compare the likelihood of \(s_y\) when paired with the original (informative) \(s_x\) versus when \(s_x\) is replaced by a non-informative variant \(\tilde{s}_x\). This transformation highlights the contribution of \(s_x\)’s semantic or structural content. Because there is inherent ambiguity in removing information from text, we employ three baselines: masked, where each token in \(s_x\) is replaced by a placeholder, preserving token count and general syntactic structure; shuffled, where words in \(s_x\) are randomly permuted, thereby disrupting the original word order while retaining lexical items; and jabberwocky, where content words are replaced with invented nonsense words, preserving an approximate syntactic scaffold.

Let \(\{(s_{x,i}, s_{y,i})\}_{i=1}^n\) be \(n\) observations of predictor–outcome text pairs. For each pair, the LLM generates a set of latent contexts \(\{c_{j;i}\}_{j=1}^{n_c}\) by conditioning on the informative \(s_x\). Within the same set of contexts \(\{c_{j;i}\}\), we compute the LLM-assessed probability of \(s_y\) when the predictor is replaced by each variant:
\[
\Pr(s_y \mid c_{j;i}, s_x^\mathrm{inf}), \quad
\Pr(s_y \mid c_{j;i}, s_x^\mathrm{mask}), \quad
\Pr(s_y \mid c_{j;i}, s_x^\mathrm{shuffle}), \quad
\Pr(s_y \mid c_{j;i}, s_x^\mathrm{jabberwocky}).
\]

We define a clause function
\[
Cl\bigl(s_y, s_x^\nu, c_{j;i}, \mM\bigr)
\;=\;
\log \Pr(s_y \mid s_x^\nu, c_{j;i}, \mM),
\]
where \(\nu \in \{\mathrm{inf}, \mathrm{mask}, \mathrm{shuffle}, \mathrm{jabberwocky}\}\) indexes the variant of \(s_x\) used. By holding the context fixed across variants, the differences in \(Cl\) between \(s_x^\mathrm{inf}\) and each noninformative version isolate the impact of the textual features that have been modified. While not a formal causal mediation effect, as any manipulation is model-based, an analogue of the indirect effect can be expressed as the ratio
\[
\frac{\Pr(s_y \mid s_x^\mathrm{inf}, c_{j;i})}
      {\Pr\bigl(s_y \mid \tilde{s}_x, c_{j;i}\bigr)},
\]
where \(\tilde{s}_x\) may represent any of the non-informative variants. Comparing these ratios for the different baselines quantifies the extent to which semantic content, word order, or lexical information contribute to the model’s prediction of \(s_y\). This setup follows the general structure of mediation analysis \citep{pearl2012causal,imai2010general}, where the generated contexts serve as mediators linking the predictor to the outcome. By systematically altering the predictor while keeping the latent context fixed, we evaluate how semantic and syntactic modifications alter the probability of the outcome string.

To formally aggregate these effects, we implement a regression model of the form
\begin{align}
\log \Pr\bigl(&s_{y,i} \mid s_{x,i}^\mathrm{inf},  c_{j;i}\bigr) 
=  \notag \\
&\beta_0 
+ \beta_1\log \Pr\bigl(s_{y,i} \mid s_{x,i}^\mathrm{shuffle},  c_{j;i}\bigr) 
+ \beta_2\log \Pr\bigl(s_{y,i} \mid s_{x,i}^\mathrm{jabberwocky},  c_{j;i}\bigr) +  \\&\quad \beta_3\log \Pr\bigl(s_{y,i} \mid s_{x,i}^\mathrm{mask}, c_{j;i}\bigr) 
+   \mathbf x_i^\top \gamma + z_i^\top b
+ \varepsilon_i.
\end{align}\label{eq:stringreg}
where $\mathbf x_i$ are string level covariates and $\mathbf z_i$ are string level random-effects.

\begin{table}[tb!]
\centering
\resizebox{\textwidth}{!}{%
\begin{tabular}{rrr}
\toprule
\textbf{Predictor Strings} & \textbf{Informative Outcome Strings} & \textbf{Placebo Outcome Strings} \\
\midrule
I love this movie & The acting was great & The capital of Pakistan is Islamabad \\
The plot was predictable & I loved the book as well & The capital of Brazil is Brasilia \\
Amazing cinematography & The storyline was captivating & The capital of France is Paris \\
\bottomrule
\end{tabular}\label{table:regressionsentences}
}
\caption{\textbf{Sample Predictor–Outcome Pairs.}}\label{table:regressionsentences}
\end{table}

\subsection{Empirical Demonstration}

For a simple illustration,  we generated a dataset consisting of 10 text-based predictor–outcome pairs, where predictors are movie-related phrases and outcomes either correspond to a semantically related movie review statement or a randomly assigned geographical fact (placebo condition). Table \ref{table:regressionsentences} presents the first three strings in each group. 

From each predictor, we generate the three different baselines:  the syntactic baseline, generated by scrambling word order while preserving vocabulary; the lexical baseline, which replaces each word or token with a non-informative mask token; and the semantic baseline, which uses a jabberwocky transformation to replace content words with nonsensical placeholders while retaining syntactic form. For example, given the predictor “The soundtrack was mesmerizing,” the shuffled version might be “was mesmerizing the soundtrack,” the masked version “<mask> <mask> <mask> <mask>,” and the jabberwocky version “The ziggflorp was blarptastic.”  We then estimate the model in Equation \ref{eq:stringreg}, including random effects for each string.  

The regression results, shown in Table \ref{table:regressionresults}, highlight the distinction between syntactic and semantic contributions to LLM-based text prediction. In both the informative and placebo settings, the shuffled predictor remains significant, confirming that syntactic coherence alone can contribute to predictability—if a sentence is grammatically well-formed, another well-formed sentence is more likely to follow, even in cases where the meaning is unrelated. However, a key distinction emerges with the jabberwocky transformation: in the informative case, where the outcome remains within the same semantic domain as the predictor, the jabberwocky effect is strongly positive and significant, suggesting that even when lexical content is removed, the syntactic structure of the predictor still carries meaningful information. In contrast, in the placebo setting, where the outcome is independent of the predictor’s meaning, the jabberwocky transformation ceases to have any significant effect. This divergence indicates that the informative regression captures a genuine semantic relationship, whereas the placebo regression isolates only syntactic structure. The masked condition, which removes all lexical content while preserving the general length and shape of the sentence, remains significant in both cases, but with a much larger coefficient in the placebo setting, suggesting that when no real information is available, the model defaults to prior expectations rather than making meaningful inferences. Together, these findings confirm that while syntactic structure alone can generate statistical dependencies between texts, only in the informative regression does the model identify an effect that is truly semantic.

\begin{table}[tb!] \centering 
\begin{tabular}{@{\extracolsep{5pt}}lcc} 
\\[-1.8ex]\hline 
\hline \\[-1.8ex] 
 & Informative & Placebo \\ 
\\[-1.8ex] & (1) & (2)\\ 
\hline \\[-1.8ex] 
 Semantic Baseline & 0.369$^{***}$ & 0.287$^{***}$ \\ 
  & (0.054) & (0.068) \\ 
  & & \\ 
 Syntactic Baseline & 0.478$^{***}$ & 0.018 \\ 
  & (0.114) & (0.122) \\ 
  & & \\ 
 Lexical Baseline & 0.359$^{***}$ & 0.687$^{***}$ \\ 
  & (0.122) & (0.109) \\ 
  & & \\ 
 Constant & 2.918$^{***}$ & 0.745 \\ 
  & (1.130) & (1.037) \\ 
  & & \\ 
\hline \\[-1.8ex] 
Fixed Effects & No & No \\ 
\hline \\[-1.8ex] 
Observations & 100 & 100 \\ 
\hline 
\hline \\[-1.8ex] 
\textit{Note:}  & \multicolumn{2}{r}{$^{*}$p$<$0.1; $^{**}$p$<$0.05; $^{***}$p$<$0.01} \\ 
 & \multicolumn{2}{r}{Random effects included at observation level.} \\ 
\hline 
\hline \\[-1.8ex] 
\end{tabular}   \caption{\textbf{Regression results using synthetic data from Table \ref{table:regressionsentences}.} The regression model distinguishes between semantic and syntactic dimensions of prediction. In the informative and placebo settings, the syntactic baseline predictor remains significant: one syntactically valid sentence is likely to follow another.  In the placebo setting, the semantic baseline does not predict the outcome, as the sentences are unrelated.  } 
  \label{table:regressionresults} 
\end{table}

\section{Replication: Estimating the Effect of a Treatment Given Text-Valued Outcomes}
\label{sec:stm_treatmenteffect}

To evaluate the effectiveness of context augmentation for testing text-based treatment effects, we replicate Experiment 3 from \citet{egami2022causal}. In that study, respondents are randomly assigned to one of two prompts, each describing a 28-year-old man who illegally entered the U.S. with one of two strings included. The treatment condition includes the sentence "The man has two prior prison sentences (one for a violent crime) and has previously been deported," while the control condition includes "The man has no prior criminal history and has never been imprisoned." Participants respond to the question: \textit{"Should this person go to jail?"} with free-text justifications. After removing observations with no open-ended response, we have $n=1034$ with $518$ treated and $516$ control.

In the original study, the authors split the data into estimation and inference sets, then fit a Structural Topic Model \citep[STM;][]{roberts2014structural} to the responses in the estimation set, using the treatment variable as a predictor for topic prevalence. Topics identified in the estimation set were then applied to the held-out inference set, with inference performed on 11 selected topics.

We apply our context augmentation method to the same data. For each response, we construct a composite prompt that includes the assigned treatment or control vignette, followed by the respondent's justification and a question-answer format:
\begin{quote}
\texttt{You were asked, “Should this person go to jail?” You replied: [response]. Why did you say this?}
\end{quote}
This prompt is passed to the model to generate contexts, which are then evaluated using our context augmentation approach.

We compare context augmentation and STM on two dimensions: power and interpretability. Using Hotelling's $T^2$ statistic, STM yields a test statistic of $18.62$ on 10 degrees of freedom, whereas context augmentation yields $858.49$ on 1 degree of freedom.  Adjusted for differences in degrees of freedom via the Wilson-Hilferty approximation, context augmentation achieves a $z$-score of $37.24$, compared to STM’s $3.83$, indicating greater statistical precision.

This improved performance stems from context augmentation leveraging additional external linguistic knowledge from a pre-trained LLM, enabling fine-grained analysis at the context level. STM, in contrast, provides a thematic summary by aggregating word co-occurrences, offering insights into broader themes rather than specific context-level contributions.

We present leave-one-out context analyses  in Table~\ref{tab:context_effects_5} in Appendix~\ref{app:context}. Contexts are divided into whether adding the context amplifies, dampens, or has no impact on the sample $t$-statistic. Positive values denote a context that increases the difference between treatment and control, while negative values decrease it.  We find positive values here correspond with contexts that reference prison as a consequence of the crime (``Yes$\ldots$should go to prison$\ldots$prior violent offender'') or, in one case, advocates for leniency, strengthening the control condition (``I don't believe in borders … fine and let him go'').  The negative values correspond with unconditional framings that applying[sic] to immigrants broadly regardless of the crime (``too many chances $\ldots$any illegal deed'', ``Probably$\ldots$we should help him$\ldots$we are a land of immigrants.'')  The contexts with no effect include signals of ambivalence or contradiction (``Deserves a chance$\ldots$ but broke the law'', ``I am torn'', ``The more illegal immigrants entering the country the more money is needed to incarcerate the many illegal immigrants'').	

We next do the same, but at the level of the observed strings.  Results are in  Table~\ref{tab:combined_loo_string} in Appendix~\ref{app:context}.  The table is constructed identically, reporting the extent to which a string-level $t$-statistics is above, at, or below the sample average.  Strings that amplify the $t$-statistics deal explicitly with recidivism or argues against incarceration above the baseline effect (``He’s a repeat offender who didn’t learn'', ``Prison doesn’t make sense; maybe just deport'').  Neutral strings offer ambivalence (``Violent records matter but I see both sides'') or facts common to both conditions (``Entered illegally; knew what he was doing''). Negative strings downplay violent priors (``Why pay for prison? Deport—it’s cheaper'') or focuses only on the immigration offense and minimizes violent priors (``I do not believe he should be sent to prison$\ldots$ do not think we we [sic] should add to overcrowded prisons'').

Note that in estimating an effect for each observed string, we are utilizing string-level estimates under each treatment condition: we estimate the mean log-probability of each string under both its observed and counterfactual treatment condition, using contexts as auxiliary information.  This allows for string-level predictions and analyses, and leaves a path open to future methods targeting causal estimands.

\begin{table}[tb!]
\centering
\resizebox{\textwidth}{!}{%
\begin{tabular}{lrrrrr}
\toprule
 & $n_c = 5$ & $n_c = 1$ & $n_c = 0.5$ & $n_c = 0.25$ & $n_c = 0.1$ \\
\midrule
Number of Generations       & 5170     & 1034     & 517     & 259   & 103   \\
Generation Time             &  6m 58s  &  1m 23s &  0m 42s &  0m 21s & 0m 8s \\
Number of Inference Evaluations        & 5,984,180 & 1,196,836 & 598,418 & 299,209 & 119,684 \\
Inference Time              & 33h 33m 53s & 6h 42m 47s & 3h 21m 23s & 1h 40m 41s & 40m 17s \\
Total Time                  & 33h 40m 51s & 6h 44m 10s & 3h 22m 5s & 1h 41m 2s & 40m 25s \\
Cross-fit $t$-statistic      & 29.4     & 26.2     & 23.7    & 19.9    & 14.1    \\
\bottomrule
\end{tabular}
}
\caption{\textbf{Computation Times.} Computation time and resulting $t$-statistic for varying context budgets $n_c$, with  $n = 1034$ strings.  Computational time is determined by the number of inference evaluations in the two-sample problem, comparing all strings over all contexts.  The first row has the number of contexts per string, with values below one of $0.5, 0.25$ indicating one context sampled  from every other and fourth string, respectively.  Inference calculations are of order $n^2 \times n_c$.  The cross-fitting takes $<3$ seconds and is not included in the calculations.  The results are still statistically significant even with few contexts generated. }\label{tab:timings}\end{table}

\subsection{Implementation and Computational Considerations} Context augmentation, while computationally intensive, is possible on the computational and time constraints used for normal research.  While the problem of selecting contexts bears some similarity to the problem of knot selection in nonparametric regression  \citep[e.g.,][]{Agarwal1980, Gu2013,Helwig2016}, a formal statement of how $n_c$ should relate to $n$ is beyond the scope of this paper. However, we can illustrate how different choices of $n_c$ affect computation time and the resulting test statistics (see Table \ref{tab:timings}).  All analyses were done using a FLAN-T5-XXL 11bn parameter model \citep{flan_T5_XXL} using 16-bit quantization on a single H100.  Computational time is determined by the number of inference evaluations in the two-sample problem, comparing all strings over all contexts.  The first row has the number of contexts per string, $n_c$, with values below one of $0.5, 0.25$ indicating one context sampled  from every other and fourth string, respectively.  

Inference calculations are of order $n^2 \times n_c$, with each of $n$ strings evaluated against contexts generated off all other strings, for $(n-1)\times n_c$ total contexts. Cross-fitting is done by subsampling from the pre-computed matrix of log-probabilities.  The cross-fitting takes $<10$ seconds and is not included in the table calculations.  The results are still statistically significant, $t=14.1$, even with few contexts generated, suggesting that smaller models can be built and, time permitting, longer runs done. 

Engineering the prompts in order to generate contexts itself relies on a rapidly evolving set of tools \citep[see, e.g.][]{chen2025unleashing,sahoo2024systematic}.  From the view of context augmentation, prompt engineering is a normal part of model-building.  While the exact form is new, the concepts of transparency, pre-registering when possible, and sensitivity analyses carry through. In this work, to minimize the ``researcher-degrees-of-freedom'' \citep{gelman_loken_2014_crisis}, we restrict ourselves to simple and straightforward prompts, as we report.  We selected the FLAN model as, after examining a subset of contexts, this was the smallest models that gave reasonable contexts.  Commercial models are also a reliable option, at the loss of replicability and privacy.  Balancing these against computational budget and time constraints will necessarily vary from project to project.

We turn next to a setting from psychology where outcomes and predictors are both textual, which is a new area of use in statistical inference.


\section{Regression Analysis: Interactive Alignment and Integrative Repair}

To illustrate the utility of context augmentation in modelling textual interactions, we apply our method to dialogue data from the DeliData corpus, a resource comprising multi-party deliberations designed explicitly for studying group decision-making dynamics \citep{karadzhov2023delidata}. Our analytical framework relies on the foundational psychological theories of Pickering and Garrod \citep{pickering2004toward,garrod2009joint, pickering2013integrated}, who characterize dialogue as a joint activity involving tightly coupled processes of language production, comprehension, and prediction. Central to their theory is the concept of \emph{interactive alignment}, wherein interlocutors' linguistic representations spontaneously synchronize across lexical, syntactic, and semantic dimensions. Such alignment enables efficient communication by minimizing cognitive demands, yet dialogues frequently encounter moments of misalignment, necessitating \emph{integrative repair}. Such repair manifests through reliance on predictable linguistic structures, facilitating rapid restoration of common ground \citep{pickering2004toward}.

Empirical studies of deliberative dialogue across diverse contexts highlight the importance of alignment and repair mechanisms. Research on police encounters \citep{rho2023escalated}, gender dynamics in deliberative settings \citep{mendelberg2014gender}, and semantic evolution in political discourse \citep{rodr:spir:2023} underscores how linguistic alignment shapes conversational trajectories. Yet, these studies often rely on human-coded interpretations of text, little or no attention paid to uncertainty estimation, or embeddings models that cannot capture syntactic attributes of text. In contrast, our context augmentation approach explicitly integrates linguistic variability into a formal inferential framework by leveraging large language model (LLM)-generated contexts, thus addressing methodological gaps noted by \citet{lemens2023semantics}.

Using our context augmentation regression, we find evidence of integrative repair within the DeliData,  manifested as increased reliance on predictable syntactic structures when confronted with a shift either into or out of consensus within the pair. Specifically, we consider utterance transitions within dialogues, modelling each subsequent utterance's log-probability conditional on its predecessor. To disentangle syntactic, semantic, and lexical contributions, we use baseline transformations: a \emph{syntactic baseline} (shuffled predictor), a \emph{semantic baseline} (jabberwocky predictor), and a \emph{lexical baseline} (masked predictor), isolating separate linguistic dimensions. These transformations isolate linguistic dimensions while preserving core textual properties.

The utterance, $s_{y;i}$ follows $s_{x;i}$ in the dialogue. We will also incorporate moderators that are measured between the $s_{x,i}$ statement and the one immediately previous.  After fitting a baseline with no moderator, we will consider three moderators:  a time variable from 0 to 1 of how much of the dialogue is complete; a consensus indicator for whether both speakers agreed prior to the predictor utterance; and a shift indicator, for whether consensus was either established or lost immediately prior to the predictor utterance.   We take this shift indicator as a measure of integrative repair, as there is a shift in understanding in the previous period, and we expect alignment afterwards.

Using random effects for string ($ u_{obs(i)}$), a dialogue-level random effect and time trend ($v_{dialogue(i)}+  time \times v_{dialogue(i)}$) and an effect for each speaker $w_{speaker(i)}$, we estimate the following model.
\begin{align}
\log &\Pr(s_{y,i} \mid s_{x,i}^{inf}, c_{j;i}) \nonumber \\
&=\beta_0 + \beta_1 \log \Pr(s_{y,i}\mid s_{x,i}^{shuffle}, c_{j;i})
+\beta_2 \log \Pr(s_{y,i}\mid s_{x,i}^{jabberwocky}, c_{j;i})
+\beta_3 \log \Pr(s_{y,i}\mid s_{x,i}^{mask}, c_{j;i}) \nonumber\\
&\quad + \gamma_1 \mathit{consensus}_i + \gamma_2 \mathit{shift}_i + \gamma_3 \mathit{moderator}_{ij}   \nonumber\\
&\quad + \delta_1(\log \Pr(s_{y,i}\mid s_{x,i}^{shuffle}, c_{j;i}) \times \mathit{moderator}_{ij}) + \delta_2(\log \Pr(s_{y,i}\mid s_{x,i}^{jabberwocky}, c_{j;i}) \times \mathit{moderator}_i) \nonumber\\
&\quad + \delta_3(\log \Pr(s_{y,i}\mid s_{x,i}^{mask}, c_{j;i}) \times \mathit{shift}_i) \nonumber\\
&\quad + u_{obs(i)} +   v_{dialogue(i)}+  time \times v_{dialogue(i)} + w_{speaker(i)} + \epsilon_i.
\end{align}

The results, summarized in Table~\ref{table:regressionresults}, confirm several important theoretical predictions. First, all three baselines significantly predict future utterances, underscoring the combined contribution of syntactic, semantic, and lexical structures to linguistic alignment. Second, there is no significant linear trend across dialogue exchanges nor a simple direct effect of consensus state. Most crucially, however, the interaction between consensus shifts and the syntactic baseline is strongly positive and statistically significant ($p<0.01$). This robust interaction indicates that shifts between consensus and dissensus systematically amplify reliance on predictable syntactic forms. We distinguish between shifting into agreement and into disagreement in Table \ref{tab:regressionresultsagree} in Appendix \ref{app:regression} ; the results are qualitatively similar, all consistent with the integrative repair hypothesis of \citet{pickering2004toward}.  By contrast, semantic and lexical baselines reveal negligible interaction effects, indicating that integrative repair primarily involves syntactic adjustment rather than semantic or lexical modification. The asymmetry underscores that syntactic structure provides a crucial scaffolding mechanism facilitating rapid realignment during conversational disruptions. These findings are robust across model specifications and clustering adjustments at the dialogue and speaker level.

\begin{table}[!tb] \centering 
  \resizebox{.9\textwidth}{!}{%
\begin{tabular}{@{\extracolsep{5pt}}rcccc} 
\\[-1.8ex]\hline 
\hline \\[-1.8ex] 
  \multicolumn{5}{c}{\textit{Dependent variable: Log Probability of Utterance given Previous Utterance }} \\ 
\\\multicolumn{1}{r}{\textbf{Moderator:} }& None & Time & Consensus & Shift \\ 
\hline \\[-1.8ex] 
Syntactic Baseline & 0.200$^{***}$ & 0.200$^{***}$ & 0.195$^{***}$ & 0.189$^{***}$ \\ 
  & (0.006) & (0.006) & (0.006) & (0.006)  \\ 
Semantic Baseline & 0.158$^{***}$ & 0.158$^{***}$ & 0.161$^{***}$ & 0.160$^{***}$  \\ 
  & (0.006) & (0.006) & (0.007) & (0.007)  \\ 
Lexical Baseline & 0.425$^{***}$ & 0.425$^{***}$ & 0.417$^{***}$ & 0.423$^{***}$ \\ 
  & (0.015) & (0.015) & (0.017) & (0.018) \\ 
  \hdashline Moderator &  & 0.108 & 0.514 & 0.390   \\ 
  &  & (0.550) & (0.393) & (0.364)   \\ 
Syntactic Baseline x Moderator &  & $-$0.003 & 0.021 & 0.045$^{***}$   \\ 
  &  & (0.019) & (0.014) & (0.013)   \\ 
Semantic Baseline x Moderator &  & 0.0003 & $-$0.016 & $-$0.008   \\ 
  &  & (0.021) & (0.015) & (0.014)   \\ 
Lexical Baseline x Moderator &  & 0.001 & 0.040 & 0.006  \\ 
  &  & (0.052) & (0.037) & (0.034)   \\ 
  \hdashline
 Constant & $-$1.391$^{***}$ & $-$1.387$^{***}$ & $-$1.504$^{***}$ & $-$1.498$^{***}$  \\ 
  & (0.162) & (0.164) & (0.183) & (0.189)  \\ 
\hline \\[-1.8ex] 
Observations & 7,390 & 7,390 & 7,390 & 7,390  \\ 
\hline 
\hline \\[-1.8ex] 
\textit{Note  }  & \multicolumn{4}{r}{$^{*}$p$<$0.1; $^{**}$p$<$0.05; $^{***}$p$<$0.01} \\ 
  \multicolumn{5}{r}{Random effects included for dialogue, speaker, and utterance.} \\ 
\end{tabular} 
}  
\caption{\textbf{Regression results from dialogic data.} }
  \label{tab:regressionresults} 
\end{table} 

Methodologically, our analysis offers significant innovations over existing approaches to text-based inference. By explicitly modelling the latent variability in LLM-generated contexts, context augmentation preserves  statistical properties (asymptotic normality, unbiasedness) while accounting for linguistic complexity. Unlike fixed embedding methods \citep{rodr:spir:2023} or purely descriptive semantic analyses \citep{lemens2023semantics}, our approach provides a fully probabilistic framework accommodating textual uncertainty. This innovation represents a major advancement in bridging modern language modelling with classical empirical process theory.

\section{Discussion and Conclusion}

This paper develops a framework for integrating large language models (LLMs) with statistical inference via context augmentation, a method that utilizes model-generated contexts  to provide uncertainty estimates and incorporate auxiliary information. With these model-generated contexts, we can estimate relationships between different sets of strings in a way that allows for standard frequentist inference, including hypothesis testing, regression modelling, and two-sample comparisons. 

The procedure rests on a standard set of identification assumptions and returns estimators that are consistent and asymptotically normal. In empirical applications, we show that the two-sample test can detect differences between groups of texts using a cross-fit $t$-statistic, and that the regression framework can capture syntactic and semantic effects using a linear model. The approach makes it possible to conduct inference directly on text data using conventional statistical tools.  The same framework admits several potential extensions. The influence function representation and identification assumptions connect naturally to the causal inference and semiparametric literature.  Context augmentation can be extended to text-based treatment effects, and the aggregation step could also be extended to account for more advanced estimation techniques.

We note two remaining issues. First, it is not clear whether the tail-conditions or approximation error rates needed for valid inference obtain.  In settings with erratic or heavy-tailed behaviour, rank-based and other robust estimators, e.g., Wilcoxon-type tests or Hodges–Lehmann estimators, can still return valid asymptotics. Our pivotality result serves as a low-cost improvement on the split-sample approach.  Second, from a practical standpoint, LLMs introduce tuning complexity, including generation temperature, top-$k$ sampling, padding, sequence length, and so on.   We used the same model to generate and score contexts, for replicability, fine control, and privacy concerns.  That said, commercial services can offer the ability to implement our framework at a reasonable cost; any model can generate contexts and many popular commercial models return the log-probability score.  Various hybrid approaches, where contexts are generated on one model and scored on another, are also a possibility.  Note that our regression model, but not the two-sample test, requires a mask token that is only available to bidirectional models.  As with earlier computational methods, we expect established conventions and best practices to emerge. We recommend reporting robustness to generation settings, via ablation or repeated sampling, as a routine part of applied work.

Context augmentation turns LLMs from predictive engines into tools for statistical inference. The method offers one path forward for integrating natural language data into the broader framework of statistical modelling. As LLMs improve, we expect that context-based inference will support more applications in text-based causal analysis, evaluation of linguistic interventions, and dialogue modelling.

\bibliographystyle{abbrvnat}
\bibliography{bibliography}

\singlespacing{
\textbf{Word count:}

\emph{$\quad$Abstract: 183}

\emph{$\quad$Body:  11,486}
}

\newpage
\appendix
\section*{ Appendix }

\counterwithin{table}{section}  
\section{Proofs}\label{app:proofs}

\subsection{Identification} \label{proof:identification}.\begin{proof}
Overlap allows the auxiliary event-context distribution to inform the distribution of strings under each event. Weak Ignorability ensures that the event-contexts isolate the effect of the event itself when evaluated with the clause function. The result follows directly from injectivity.
\end{proof}


\subsection{Influence Function Decomposition} \label{app:vonmises}.  

Assumption 1 allows an estimating equation that is a sample average over independently distributed strings and their associated contexts. Assumption 2 guarantees that the estimating problem is well-posed. Assumptions 3 and 4 provide the regularity conditions required for the functional delta method \citep[see][Thm. 20.8]{vand:1998}. Assumption 5 ensures a uniform approximation in a neighborhood of the true value. Together, these conditions yield the reported von Mises expansion.

\subsection{Limit Theorem}\label{app:CLT}.  

 The Lyapunov condition ensures asymptotic normality of the leading term of the von Mises expansion from the previous result. The uniform continuous mapping theorem and Slutsky’s method justify the limiting distribution. The result is directly comparable to the central limit theorem for regular Z-estimators with nuisance functions \citep[see][Thm.~25.54 and 25.57]{vand:1998}.

\subsection{Pivotality and Rate Result}\label{app:pivotresult}

The repeated–cross–fit estimator is  a symmetric
U-statistic of order two in the sample indices; see Appendix \ref{app:estimation} for implementation details.   \citet{EscancianoTerschuur2022} model a single cross-fit as a U-statistic, while  \citet{WangLindsay2014} give an unbiased variance estimate for the repeated cross-fit estimate \citep[see][for general theory]{politis1999subsampling}.  In our setting, the influence function
\(\varphi\) lies in \(L_{2}(P)\) and satisfies a \(2+\varepsilon\) Lyapunov moment
condition, this average has finite variance and is automatically tight.

The higher-order efficiency result follows the logic of
\citet{hall1992bootstrap}.  Under Cramér’s condition and a
\(4+\varepsilon\) bound on the moments of \( \psi_{\theta, \eta}\), a second–order Edgeworth expansion exists for the estimator \citep[see][]{bhattacharya1976normal}.
Pivotality removes the leading \(r_{n}^{-1}\) term in both the pivot and the
estimator, so their difference is \(o_{p}(r_{n}^{-2})\).  Consequently we can
relax the required approximation rate from \(n^{-1/4}\) to \(n^{-1/8}\).

Next, in our von~Mises decomposition the only non-parametric remainder
is the self-referential bias, which factors into
\[
  \underbrace{\|\widehat F - F\|_{\infty}}_{\text{context error}}
  \quad\times\quad
  \underbrace{\frac1M\sum_{m=1}^{M} R^{(m)}}_{\substack{\text{self-referential}\\\text{remainder}}},
\]
where \(R^{(m)}\) are i.i.d.\ mean-zero with finite variance by the argument
above.  By the Dvoretzky–Kiefer–Wolfowitz bound,
\(
\|\widehat F - F\|_{\infty}=O_{p}(\sqrt{\ln n_{c}/n_{c}})
\),
while the Kolmogorov LIL gives
\(
M^{-1}\!\sum_{m=1}^{M} R^{(m)}=O_{p}(\sqrt{\ln\!\ln M\,/M})
\).
Hence the bias is
\(
O_{p}\bigl(\sqrt{\tfrac{\ln n_{c}}{n_{c}}\,\tfrac{\ln\!\ln M}{M}}\bigr)
\),
and solving \(o(n^{-\gamma})\) (absorbing logs into an \(\varepsilon>0\)) yields
\(
(n_{c}M)^{1+\varepsilon}\gg n^{2\gamma},
\)
as reported in the text.

\section{Estimation} \label{app:estimation}

To avoid self-referential bias, we implement a repeated cross-fitting procedure. Let \( \mathcal{I}^1 \) and \( \mathcal{I}^2 \) denote a random partition of the sample into two folds. For each string \( s_i \in \mathcal{I}^1 \), we evaluate it on out-of-fold contexts generated by strings in \( \mathcal{I}^2 \), and vice versa. Specifically, define the string-level statistic as
\[
\widehat{Str}_i^{\mathcal{E};\mathcal{I}^2} = 
\begin{bmatrix}
\frac{1}{|C^A|} \sum_{c \in C^A} \log \Pr(s_i \mid c) \\
\frac{1}{|C^B|} \sum_{c \in C^B} \log \Pr(s_i \mid c)
\end{bmatrix}, \quad \text{for } s_i \in \mathcal{I}^1,
\]
where \( C^G \) denotes the set of contexts generated from group \( G \)'s strings in fold \( \mathcal{I}^2 \). The difference-in-means estimator for group A is then
\[
\widehat{\theta}^{A; \mathcal{I}^2} = \frac{1}{n_{A1}} \sum_{i \in \mathcal{I}^1 \cap A} \left( \widehat{Str}_{i,A}^{\mathcal{E};\mathcal{I}^2} - \widehat{Str}_{i,B}^{\mathcal{E};\mathcal{I}^2} \right),
\]
and similarly for group B using \( \mathcal{I}^2 \cap B \).

We then reverse the roles of the folds and average:
\[
\widehat{\theta}^{\text{avg}} = \frac{1}{2} \left( \widehat{\theta}^{A; \mathcal{I}^2} - \widehat{\theta}^{B; \mathcal{I}^2} + \widehat{\theta}^{A; \mathcal{I}^1} - \widehat{\theta}^{B; \mathcal{I}^1} \right).
\]

To account for sampling variability, we compute a t-statistic in each direction and average:
\[
\widehat{t} = \sqrt{2} \cdot \frac{1}{2} \left( t_{\mathcal{I}^1 \rightarrow \mathcal{I}^2} + t_{\mathcal{I}^2 \rightarrow \mathcal{I}^1} \right),
\]
where \( t_{\mathcal{I}^1 \rightarrow \mathcal{I}^2} \) is the t-statistic computed by evaluating strings in \( \mathcal{I}^1 \) on contexts from \( \mathcal{I}^2 \), and vice versa.

This estimator is repeated across multiple random splits, and final inference is based on the distribution of \( \widehat{t} \) across cross-fits.

\newpage

\section{Two-Sample Context Augmentation: Prompting and Evaluation}
\label{app:twosamplepseudocode}

\begin{algorithm}
\caption{Two-Sample Context Augmentation}
\begin{algorithmic}[1]
\Require String lists $S_A,S_B$; number of contexts $J$; cross-fits $R$
\Ensure  Mean cross-fit $t$, two-sided $p$
\State \textbf{Prompt settings:} see Box~\ref{box:prompts}.
\For{each string $s\in S_A\cup S_B$}
  \State (a) \textbf{Definition:}  $\texttt{def}\leftarrow$\,LLM(\texttt{"Provide a clear and concise definition of the word '\(s\)'."})
  \State (b) \textbf{Left fragment} $L_j\ (j=1{:}J)$\\
        \hspace*{1.5em}\texttt{"Using the definition '\(<\!def\!>\)', provide a sentence fragment that could logically come before the word '\(s\)' and conveys its meaning."}
  \State (c) \textbf{Right fragment} for each $L_j$\\
        \hspace*{1.5em}\texttt{"Using the definition '\(<\!def\!>\)', provide a sentence fragment that could logically come after the phrase '$<\!L_j\!>\;s$' and conveys its meaning."}
  \State \hspace*{1.5em}template$_j \gets L_j\;\mathbf{<<<STR>>>}\;R_j$
\EndFor
\State \textbf{Fill \& score:} obtain log-probabilities $\ell_{s,j}\leftarrow \log\Pr(s\mid\text{template}_j)$
\State \textbf{Repeated cross-fitting} (run $r=1{:}R$):
  \begin{enumerate}
    \item Split $S_A,S_B$ into halves $(A_1,A_2)$, $(B_1,B_2)$.
    \item Evaluate $A_2$ only on contexts from $A_1$ vs.~$B_1$ (and vice-versa).
    \item Compute two $t$-statistic, average, rescale by $\sqrt2$ to adjust for sample size.
  \end{enumerate}
\State \textbf{Return} $\displaystyle \bar t=\frac1R\sum_{r=1}^R t^{(r)}$ and $p=2\!\left[1-F_{|t|}\!\bigl(|\bar t|\bigr)\right]$
\end{algorithmic}
\end{algorithm}

\bigskip
\noindent\fcolorbox{black}{lightgray}{%
\parbox{0.95\linewidth}{%
\textbf{Box 1: Prompt Settings}\label{box:prompts}

\medskip
\textbf{(a) Simulation (Synthetic categories)}\\
Definition prompt: \texttt{"Provide a clear and concise definition of the word '\textless word\textgreater'."}\\
Left prompt: \texttt{"Using the definition '\textless definition\textgreater', provide a sentence fragment that could logically come before the word '\textless word\textgreater' and conveys its meaning."}\\
Right prompt: \texttt{"Using the definition '\textless definition\textgreater', provide a sentence fragment that could logically come after the phrase '\textless left\textgreater\ \textless word\textgreater' and conveys its meaning."}\\
Hyper-parameters: \texttt{temperature}=0.8,\; \texttt{top\_k}=50,\; \texttt{do\_sample}=True

\bigskip
\textbf{(b) Replication (Egami et al.)}\\
\emph{Treatment base vignette}\\
\texttt{"A 28-year-old single man, a citizen of another country, was convicted of illegally entering the United States. Prior to this offense, he had served two previous prison sentences each more than a year. One of these previous sentences was for a violent crime and he had been deported back to his home country."}\\[2pt]
\emph{Control base vignette}\\
\texttt{"A 28-year-old single man, a citizen of another country, was convicted of illegally entering the United States. Prior to this offense, he had never been imprisoned before."}\\[4pt]
Definition prompt: \texttt{"\textless base\_prompt\textgreater\ You were asked, 'Should this person go to jail.' You replied: '\textless word\textgreater'. Why did you say this?"}\\
Left prompt: \texttt{"Using the definition '\textless definition\textgreater', provide a sentence fragment before the word '\textless word\textgreater'."}\\
Right prompt: \texttt{"Using the definition '\textless definition\textgreater', provide a sentence fragment after '\textless left\textgreater\ \textless word\textgreater'."}\\
Hyper-parameters: \texttt{temperature}=0.8,\; \texttt{top\_k}=50,\; \texttt{do\_sample}=True
}%
}

\section{Text-on-Text Regression: Generation and Evaluation}
\label{app:regressionpseudocode}

\begin{algorithm}
\caption{Context-Mediated Text Regression}
\begin{algorithmic}[1]
\Require Predictor strings $\{s_{x,i}\}$, outcome strings $\{s_{y,i}\}$, contexts $J$
\Ensure  Regression coefficients $\beta_{1{:}3}$ with clustered SEs
\For{each observation $i$}
  \State \textbf{Context generation:} $c_{i,1{:}J}\leftarrow\text{LLM}(\text{context prompt}\,|\,s_{x,i})$
  \State \textbf{Build predictor variants}: informative, masked, shuffled, jabberwocky
  \For{$j=1{:}J$}  \Comment{$Cl$ evaluation}
     \State Two orders:  
            $c^{PX}_{i,j}= \text{variant}+c_{i,j}+s_{y,i}$,\,
            $c^{XP}_{i,j}= s_{y,i}+c_{i,j}+\text{variant}$
     \State $\ell^{PX}_{i,j}\gets\text{LLM}(\text{eval prompt},c^{PX}_{i,j})$,\;
            $\ell^{XP}_{i,j}\gets\text{LLM}(\text{eval prompt},c^{XP}_{i,j})$
     \State $\ell_{i,j}\gets\max(\ell^{PX}_{i,j},\ell^{XP}_{i,j})$
  \EndFor
  \State $\text{Str}_i=\mathcal{A}(\{\ell_{i,j}\}_{j=1}^J)$   \Comment{mean over contexts}
\EndFor
\State \textbf{Regression ($T$ operator)}:\\
\hspace*{1em}$\log\Pr(s_y|s_x^{\text{inf}})=
\beta_0+\beta_1\log\Pr(\cdot|s_x^{\text{shuf}})+
\beta_2\log\Pr(\cdot|s_x^{\text{jab}})+
\beta_3\log\Pr(\cdot|s_x^{\text{mask}})+\ldots+\varepsilon$
\end{algorithmic}
\end{algorithm}

\bigskip
\noindent\fcolorbox{black}{lightgray}{%
\parbox{0.95\linewidth}{%
\textbf{Box 2: Regression‐Specific Prompts}\label{box:jab}

\medskip
\textbf{Masked baseline}: replace every token in $s_x$ with \texttt{\textless mask\textgreater}.\\
\textbf{Shuffled baseline}: randomly permute words in $s_x$.\\
\textbf{Jabberwocky (multi‐shot) first example}\\
\texttt{Input: 'The quick brown fox jumps over the lazy dog.'}\\
\texttt{Step 1: Identify nouns (fox, dog), verbs (jumps), adjectives (quick, brown, lazy).}\\
\texttt{Step 2: Fix function words (articles/prepositions).}\\
\texttt{Step 3: Replace content words with nonsense: nouns (blob, zink); verb (scroop); adjectives (froopy, blorn, flark).}\\
\texttt{Output: The froopy blorn blob scroops over the flark zink.}\\[4pt]
\textbf{Context prompt (Synthetic regression)}:\\
\texttt{"Reflect on '\textless predictor\textgreater'. Assume a paragraph contains '\textless predictor\textgreater'. Provide a detailed reflection that captures its essence."}\\[4pt]
\textbf{Context prompt (Delidata)}:\\
\texttt{"Reflect on the following dialogue cue: '\textless predictor\textgreater'. Imagine this statement is part of a two-person deliberative discussion where the participants are solving a card selection task. Generate a detailed background reflection that captures the underlying strategic reasoning, deliberative dynamics, and potential influences of group consensus. Focus on how the predictor might reveal the participants' thought processes and decision-making strategies."}\\[4pt]
\textbf{Evaluation prompt}: \texttt{"Score the naturalness and coherence of the following sentence:"}\\
Generation hyper-parameters: \texttt{temperature}=1.0 (jabberwocky 1.2), \texttt{top\_k}=50, \texttt{do\_sample}=True
}%
}

\medskip
\noindent\textbf{Notes.}\;Both tasks share the pipeline  
$Cl\!\to\!\mathcal{A}\!\to\!Str\!\to\!T$: clause scores $\xrightarrow{\mathcal{A}}$ context means $\xrightarrow{T}$ test statistic or regression.  
Contexts are always generated \emph{only} from $s_x$; $s_y$ never appears inside a context.

\newpage 
\section{Context and String Two-Sample Analysis}\label{app:context}

\begin{table}[ht]
\centering
\resizebox{.8\textwidth}{!}{%
\begin{tabular}{lrp{15cm}}
  \hline
 & Change in $t$ &Context \\ 
  \hline
   Amplifies & -0.0266 & I think the person should be deported to his own country. I would consider him a violet[sic] offender and maybe a career criminal. We should not waste the resources of a already $<$$<$$<$STR$>$$>$$>$ I think the person should be deported to his own country. I would consider him a violet[sic] offender and maybe a career criminal. We should not waste the resources of a already \\ 
  Amplifies & -0.0268 & Yes, I believe this man should go to prison, because he is a prior violent offender. He was previously convicted and served time in prison, and was deported back to $<$$<$$<$STR$>$$>$$>$ Yes, I believe this man should go to prison, because he is a prior violent offender. He was previously convicted and served time in prison, and was deported back to \\ 
  Amplifies & -0.0268 & They never know how they voted for the President. $<$$<$$<$STR$>$$>$$>$ I don't believe in borders. Give this man a fine and let him be on his way. Unless they have outstanding warrants or are carrying narcotics, who \\ 
  Amplifies & -0.0269 & I am concerned that this person has been deported in the past and has again entered the country illegally. He was previously convicted of a violent crime which warrants a deport $<$$<$$<$STR$>$$>$$>$ I am concerned that this person has been deported in the past and has again entered the country illegally. He was previously convicted of a violent crime which warrants a deport \\ 
  Unchanged & 0.0000 & Because he clearly broke the law. $<$$<$$<$STR$>$$>$$>$ Because he clearly broke the law because he has no valid citizenship. \\ 
  Unchanged & 0.0000 & That is what the current system is. $<$$<$$<$STR$>$$>$$>$ That is what the current system is. If you put the ones you catch in prison when you catch them maybe less will want to try it. I would hope that would be the result. \\ 
  Unchanged & 0.0000 & I don't think he should be sent to jail. If he has legitimate reasons for leaving his country and coming to America (refugee), then those need to be looked $<$$<$$<$STR$>$$>$$>$ Because he deserves a chance at a better life and should not be sent back to his country \\ 
  Unchanged & -0.0000 & I am torn. On one hand I feel that if he was here illegally, then he should possibly be sent back to his place of origin. There isn' $<$$<$$<$STR$>$$>$$>$ He was already in prison for another crime. \\ 
  Unchanged & 0.0000 & He should instead be deported. $<$$<$$<$STR$>$$>$$>$ The more illegal immigrants entering the country the more money is needed to incarcerate the many illegal immigrants. \\ 
  Dampens & 0.0253 & This person is a known criminal who has been convicted of violent crimes. In addition, they are in the country illegally. They should definitely go to prison as it is likely they will $<$$<$$<$STR$>$$>$$>$ They need to be sent to prison for their crimes so it will stop them from breaking more laws \\ 
  Dampens & 0.0247 & A man who is illegal, and is caught in the act of being illegal, should be sent to prison. When released he should be deported to his country and not be allowed to ever $<$$<$$<$STR$>$$>$$>$ He should be deported to his country, and not allowed to ever come to the US again \\ 
  Dampens & 0.0246 & This person should spend time in Prison, He's already had too many chances and he didn't learn from them. Yes he should spend time in prison or stay in his own $<$$<$$<$STR$>$$>$$>$ I believe the penalty for illegal entry should be the same as any other illegal deed \\ 
  Dampens & 0.0243 & [i] Probably illegal immigrants are refugees and we should make an effort to help and not condemn him. We are in fact a land of immagrints[sic]. We need $<$$<$$<$STR$>$$>$$>$ Probably illegal immigrants are refugees and we should make an effort to help and not condemn him. We are in fact a land of immagrints. We need to set the \\ 
  Dampens & 0.0240 & I do not know the law in this regard. $<$$<$$<$STR$>$$>$$>$ i. \\ \hline
  \end{tabular}
  }\caption{\textbf{Leave-one-out context sensitivity.} Whether each context amplifies, dampens, or leaves the effect unchanged by average $\Delta t$ shift.} \label{tab:context_effects_5}
\end{table}

\begin{table}[ht]
\centering
\resizebox{.8\textwidth}{!}{%
\begin{tabular}{lrp{15cm}}
  \hline
	 & Effect & String \\ 
 \hline
 \hline
Above Average & 0.1765 & He is a repeat offender who didnt learn a lesson \\ 
  Above Average & 0.1359 & Placing the man in prison means that our country would have to pay to feed, clothe and harbor him.  Our government would send the man back to his country of origin, or if refused, back to the country that he immigrated from.  He should be a burden upon another society, not ours. \\ 
  Above Average & 0.1286 & I don't think it makes sense to put him in prison. It will simply cost us money. Possibly deport him depending on his status \\ 
  Above Average & 0.1192 & He should be processed by immigration to see if he has the ability to legally stay in the country (such as a refugee). If that's not the case, his illegal entry should be recorded and he should be returned to his home country. \\ 
  Above Average & 0.1113 & I think that the U.S. Government should imprison the person who entered the country illegally for a short period of time, such as one month, and then deport him back to his country of origin.  If he wants to return, he should try to enter legally. \\ 
  At the Average & 0.0000 & This man entered the country illegally, breaking the law.  It doesn't matter that he had never committed a crime before.  He knowingly entered the country illegally. \\ 
  At the Average & 0.0000 & He should be deported given his violent criminal history. \\ 
  At the Average & 0.0001 & I think this man should be deported to his home country.  He broke the law and must pay. \\ 
  At the Average & -0.0001 & i do not think that illegally entering the country is an offense that deserves a prison sentence. even though he had a violent crime that he committed in the past, he did not commit a violent crime in this offense. \\ 
 At the Average & 0.0002 & I understand why people try to come here illegally, but I also understand why people would not want violent criminals. Past violent criminal records should be taken into account. \\ 
  Below Average & -0.0557 & This person should be deported to the home country, not imprisoned in the US. If this person had entered the US illegally, and committed a violent crime, it would have been appropriate to send this person to a US prison.  But this person's only current crime is illegally entering the US, so deportation is the appropriate penalty. \\ 
  Below Average & -0.0559 & I do not believe he should be sent to prison on the sole basis of entering the US illegally. I would be indifferent to deportation, but do not think we we [sic] should add to overcrowded prisons unless he had committed a more serious offense. \\ 
  Below Average & -0.0566 & It is illegal to enter the country with no paperwork. But, if he spends time in prison, he should not be sent back home, as it would be a double sentence. \\ 
  Below Average & -0.0566 & if he was here illegally he should just be sent back to where he came from. \\ 
  Below Average & -0.0567 & I don't know what the standard protocol is for this, but it sounds like he should be deported again, but I'm not sure if you typically serve prison time before that happens. I have no problem with people illegally entering the country, but if you're going to commit several crimes and still continue doing it after being deported once then maybe you should be forced to leave. Also, it's important if he has US citizenship. At this point, it doesn't sound like the person does. \\  
\hline
   \end{tabular}
}
\caption{\textbf{String-level analysis.} Strings organized by those with effect estimates at/above/below the sample average.} 
\label{tab:combined_loo_string}
\end{table}

\newpage
\section{Regression Results}\label{app:regression}

\begin{table}[!htbp] \centering 
  \resizebox{.9\textwidth}{!}{%
\begin{tabular}{@{\extracolsep{5pt}}rccccc} 
\\[-1.8ex]\hline 
\hline \\[-1.8ex] 
  \multicolumn{6}{c}{\textit{Dependent variable: Log Probability of Utterance given Previous Utterance }} \\ 
\\\multicolumn{1}{r}{\textbf{Moderator:} }& None & Time & Consensus & Shift & Agree/Disagree\\ 
\hline \\[-1.8ex] 
Syntactic Baseline & 0.200$^{***}$ & 0.200$^{***}$ & 0.195$^{***}$ & 0.189$^{***}$ & 0.189$^{***}$ \\ 
  & (0.006) & (0.006) & (0.006) & (0.006) & (0.006) \\ 
Semantic Baseline & 0.158$^{***}$ & 0.158$^{***}$ & 0.161$^{***}$ & 0.160$^{***}$ & 0.160$^{***}$ \\ 
  & (0.006) & (0.006) & (0.007) & (0.007) & (0.007) \\ 
Lexical Baseline & 0.425$^{***}$ & 0.425$^{***}$ & 0.417$^{***}$ & 0.423$^{***}$ & 0.423$^{***}$ \\ 
  & (0.015) & (0.015) & (0.017) & (0.018) & (0.018) \\ 
  \hdashline Moderator &  & 0.108 & 0.514 & 0.390 &  \\ 
  &  & (0.550) & (0.393) & (0.364) &  \\ 
Syntactic Baseline x Moderator &  & $-$0.003 & 0.021 & 0.045$^{***}$ &  \\ 
  &  & (0.019) & (0.014) & (0.013) &  \\ 
Semantic Baseline x Moderator &  & 0.0003 & $-$0.016 & $-$0.008 &  \\ 
  &  & (0.021) & (0.015) & (0.014) &  \\ 
Lexical Baseline x Moderator &  & 0.001 & 0.040 & 0.006 &  \\ 
  &  & (0.052) & (0.037) & (0.034) &  \\ 
\hdashline Agree &  &  &  &  & 0.838$^{*}$ \\ 
  &  &  &  &  & (0.485) \\ 
 Disagree &  &  &  &  & $-$0.028 \\ 
  &  &  &  &  & (0.471) \\ 
Syntactic Baseline x Agree &  &  &  &  & 0.052$^{***}$ \\ 
  &  &  &  &  & (0.017) \\ 
Semantic Baseline x Agree &  &  &  &  & $-$0.014 \\ 
  &  &  &  &  & (0.019) \\ 
Lexical Baseline x Agree &  &  &  &  & 0.043 \\ 
  &  &  &  &  & (0.046) \\ 
Syntactic Baseline x Disagree &  &  &  &  & 0.037$^{**}$ \\ 
  &  &  &  &  & (0.017) \\ 
Semantic Baseline x Disagree &  &  &  &  & $-$0.003 \\ 
  &  &  &  &  & (0.018) \\ 
Lexical Baseline x Disagree &  &  &  &  & $-$0.028 \\ 
  &  &  &  &  & (0.045) \\
  \hdashline
 Constant & $-$1.391$^{***}$ & $-$1.387$^{***}$ & $-$1.504$^{***}$ & $-$1.498$^{***}$ & $-$1.498$^{***}$ \\ 
  & (0.162) & (0.164) & (0.183) & (0.189) & (0.189) \\ 
\hline \\[-1.8ex] 
Observations & 7,390 & 7,390 & 7,390 & 7,390 & 7,390 \\ 
Log Likelihood & $-$3,984.707 & $-$3,983.085 & $-$3,982.212 & $-$3,978.490 & $-$3,977.092 \\ 
Akaike Inf. Crit. & 7,989.413 & 7,994.170 & 7,992.423 & 7,984.979 & 7,990.185 \\ 
Bayesian Inf. Crit. & 8,058.492 & 8,090.880 & 8,089.133 & 8,081.690 & 8,114.527 \\ 
\hline 
\hline \\[-1.8ex] 
\textit{Note  }  & \multicolumn{5}{r}{$^{*}$p$<$0.1; $^{**}$p$<$0.05; $^{***}$p$<$0.01} \\ 
 & \multicolumn{5}{r}{Random effects included for dialogue, speaker, and utterance.} \\ 
\end{tabular} 
}  \caption{\textbf{Finer regression results from dialogic data.} This reproduces the paper from the text, but breaks the ``shift'' moderator into shifting into agreement or disagreement.  Results are qualitatively the same for each. }
  \label{tab:regressionresultsagree} 
\end{table}

\end{document}